
\documentclass[12pt, preprint]{aastex}
\usepackage{wrapfig,lipsum,booktabs}
\usepackage{booktabs,fixltx2e}
\usepackage{threeparttable}
\usepackage[caption=false]{subfig}
\usepackage{emulateapj5}

\newcommand*{\rom}[1]{\expandafter\@slowromancap\romannumeral #1@}

\newcommand{\msun}{M$_{\sun}$\,}
\newcommand{\myr}{M$_\odot$~yr$^{-1}$} 
\newcommand{\myrkpc}{M$_\odot$~yr$^{-1}$~kpc$^{-2}$} 

\newcommand{\ha}{H$\alpha$}

\newcommand{\hb}{H$\beta$}
\newcommand{\nii}{[N{\sc II}]}
\newcommand{\oiii}{[O{\sc III}]}
\newcommand{\sii}{[S{\sc II}]}
\newcommand{\hii}{H{\sc II}}
\newcommand{\loghn}{log(\nii/\ha)}
\newcommand{\kms}{km\,s$^{-1}$} 

\newcommand{\ferg}{erg s$^{-1}$ cm$^{-2}$}
\newcommand{\ergs}{erg s$^{-1}$}
\newcommand{\mjy}{mJy}

\shorttitle{Resolving Host Galaxies of z=2 QSOs}
\shortauthors{Vayner et al. 2014}

\begin{document}
\title{Providing stringent star formation rate limits of \lowercase{z}$\sim$2 QSO host galaxies at high angular resolution}
\author{Andrey Vayner\altaffilmark{1,2}, Shelley A. Wright\altaffilmark{1,2,3}, Tuan Do\altaffilmark{2,4,7}, James E. Larkin\altaffilmark{4}, Lee Armus\altaffilmark{5}, S. C. Gallagher\altaffilmark{6}}

\altaffiltext{1}{Department of Astronomy  \& Astrophysics, University of Toronto, 50 St. George street, Toronto, ON, M5S 3H4}
\altaffiltext{2}{Dunlap Institute for Astronomy \& Astrophysics, University of Toronto, 50 St. George street, Toronto, ON, M5S 3H4}
\altaffiltext{3}{Center for Astrophysics and Space Sciences, University of California, San Diego, La Jolla, CA 90037}
\altaffiltext{4}{Department of Physics and Astronomy, University of California, Los Angeles, CA 90095}
\altaffiltext{5}{Spitzer Science Center, California Institute of Technology, 1200 E. California Blvd., Pasadena, CA 91125}
\altaffiltext{6}{Department of Physics and Astronomy, The University of Western Ontario, London, ON N6A 3K7}
\altaffiltext{7}{Dunlap Fellow}

\begin{abstract}
We present integral field spectrograph (IFS) with laser guide star adaptive optics (LGS-AO) observations of z$\sim$2 quasi-stellar objects (QSOs) designed to resolve extended nebular line emission from the host galaxy. Our data was obtained with W. M. Keck and Gemini-North Observatories using OSIRIS and NIFS coupled with the LGS-AO systems, respectively. We have conducted a pilot survey of five QSOs, three observed with NIFS+AO and two observed with OSIRIS+AO at an average redshift of z=2.2. We demonstrate that the combination of AO and IFSs provides the necessary spatial and spectral resolutions required to separate QSO emission from its host. We present our technique for generating a PSF from the broad-line region of the QSO and performing PSF subtraction of the QSO emission to detect the host galaxy emission at separation of $\sim$0.2$\arcsec$ ($\sim$1.4 kpc). We detect \ha~narrow-line emission for two sources, SDSSJ1029+6510 (z$_{H\alpha}$=2.182) and SDSSJ0925+0655 (z$_{H\alpha}$=2.197), that have evidence for both star formation and extended narrow-line emission. Assuming that the majority of narrow-line \ha~emission is from star formation, we infer a star formation rate for SDSSJ1029+6510 of 78.4 \myr~originating from a compact region that is kinematically offset by 290 - 350 \kms. For SDSSJ0925+0655 we infer a star formation rate of 29 \myr~distributed over three clumps that are spatially offset by $\sim$ 7 kpc. The null detections on three of the QSOs are used to infer surface brightness limits and we find that at 1.4 kpc from the QSO the un-reddened star formation limit is $\lesssim$ 0.3 \myrkpc. If we assume typical extinction values for z=2 type-1 QSOs, the dereddened star formation rate for our null detections would be $\lesssim$ 0.6 \myrkpc. These IFS observations indicate that while the central black hole is accreting mass at 10-40$\%$ of the Eddington rate, if star formation is present in the host (1.4 - 20 kpc) it would have to occur diffusely with significant extinction and not in compact, clumpy regions.

\end{abstract}


\section{Introduction}\label{intro}
Understanding the formation and growth of supermassive black holes (SMBH) in galaxy evolution is a key problem in astrophysics. Some of the largest puzzles are the origin of M$_{bh}$-$\sigma$ relationship (\citealt{mago1998}, \citealt{geb00}, \citealt{ferra00}), the role of active galactic nuclei (AGN) feedback and its effects for quenching star formation (e.g., \citealt{scan05,barai2014}), and how to effectively transport gas to the galactic nuclei to fuel black hole growth (e.g., \citealt{thompson2005, hopkins11}). 

The majority of quasi-stellar object (QSO) host galaxy studies have concentrated on nearby systems (z$\lesssim$0.4) using optical observations with Hubble Space Telescope (HST) and ground-based facilities (e.g., \citealt{Bahcall1997, lehnhert99, ham02, hutching2002, ridgeway02, mar03, zak06, floyd10, floyd13}). The key ingredient in these observations is to achieve the high spatial resolutions necessary to disentangle bright QSO emission from that of the underlying stellar population and HII regions. A range of host galaxy parameters has been discovered, implying that QSOs are hosted by many different galaxy types with a range of simultaneous star formation activity (e.g, \citealt{benn08}), QSOs have been found in massive inactive elliptical galaxies, late-type spirals and irregulars. Even with the large range of selection effects, there is a coherent picture that luminous nearby QSOs are generally found in luminous and massive host galaxies with a range of morphologies \citep{mats14}. However, at high-redshift (z$\gtrsim$1) the picture of QSO host galaxies is less clear with only a small number of host systems observed. High redshift QSOs have been found in star forming galaxies with morphologies ranging from discs (e.g., \citealt{inskip11}) to mergers (e.g. \citealt{carniani2013,floyd13}), while some studies have shown QSOs to reside in passive, elliptical galaxies (e.g. \citealt{kotilainen2009}). 

One of the most compelling physical explanations of the co-evolution of the host galaxy and SMBH, is negative feedback from AGN energetics. There has been mounting observational evidence supporting star formation quenching via QSO/AGN activity by expelling large reservoirs of cold gas and/or heating of the gas in massive halos (\citealt{fab12}, references therein). Recent studies have found that the majority of low redshift type-2 QSOs (z$\sim$ 0.2) contain evidence of galaxy wide outflows on kpc scales with \oiii 5007\AA~emission lines \citep{harrison2014, liu13}; however their effects on star formation rates are yet to be understood. Similarly, recent integral field spectrograph (IFS) observations of \oiii~$\&$ H$\alpha$ emission in z$\sim$2 QSOs have revealed host galaxies with strong evidence of outflows, and lower star formation rates in regions with the strongest outflows (Gemini NIFS: \citealt{alexander2010}, VLT SINFONI: \citealt{cano-diaz2012}). These observations have given tantalizing clues of QSO feedback, yet there is still little known about the z$\sim$2 host galaxies (i.e, stellar mass, dynamics, metallicities), and whether they obey the present day black hole mass-galaxy scaling relations \citep{2013ApJ...764..184M}.  

As QSOs outshine their host galaxies by several orders of magnitude, studying their hosts requires a careful removal of the QSO emission, for which a good understanding of the point spread function (PSF) is required. Understanding the PSF for ground based observations is very difficult since atmospheric variations cause the PSF to change over a time span of a few seconds, making it extremely difficult to model. There have been some successful attempts to remove the bright QSO light using nearby stars as reference to detect extended emission from the host galaxy (e.g., seeing-limited:\citealt{falomo04, kotilainen2009,schramm2008} HST/AO: \citealt{falomo05}). The majority of QSO host galaxy observations have used space-based observations where the PSFs are stable for QSO removal. At low redshift (z$\lesssim$1) there have been several studies that used both artificial and stellar PSFs to remove QSO light to search for extended emission, which have allowed for several successful studies of low and intermediate redshift QSO hosts (\citealt{Bahcall1997}, \citealt{kirhakos1999}, \citealt{hutching2002}). At high-redshift (1 $\lesssim$ z $\lesssim$4), these searches have been more challenging because the angular scales of host galaxies are comparable to the PSF halo ($\sim$1\arcsec) and PSF removal is dominated by residuals, which makes it difficult to disentangle the QSO and host galaxy. The bigger difficulty comes from extracting star formation rates (SFR) and metallicities of the host galaxies, since these quantities can be easily contaminated by QSO narrow-line emission with a range of spatial and kinematic offsets ($\lesssim$ 1000 \kms;\citealt{cano-diaz2012}, \citealt{liu13}). Broad-band photometry has been used to model the stellar properties of distant host galaxies, however residual noise from PSF subtraction makes it difficult to obtain accurate magnitudes, and there are no reliable tests to distinguish stellar rest-frame optical continuum from the synchrotron emission of the central AGN.

A combination of adaptive optics (AO) and integral field spectroscopy (IFS) provides the necessary spatial and spectral resolutions required to separate QSO emission from its host. Having spectral information at each spatial location allows us to extract key information about the galaxy that an imaging survey simply cannot achieve. IFS observations provide a powerful technique to remove the bright QSO. This can be achieved by utilising unresolved emission from the QSO (i.e., broad-line emission, like H$\alpha$) to construct a pure QSO PSF image. This PSF image is normalized and then subtracted per wavelength channel in the data cube, thus leaving only narrow-line emission. If there is spatially offset narrow-line emission, this can be used directly to infer kinematics, dynamical masses (assuming virialized gas), and nebular emission diagnostics of the gas. Recently this technique was proven to be effective in resolving the host galaxy of a redshift z=1.3 QSO using SINFONI on the VLT (\citealt{inskip11}). These authors were successful at detecting the host galaxy and were able to construct a spatially resolved narrow emission line map with identified ionization mechanisms and star formation rates (100 \myr). They found that the galaxy dynamical mass and black hole mass obeyed the present-day M$_{BH}$ vs. M$_{bulge,stellar}$ relation within the current scatter. In contrast, there have been no IFS observations of high-z QSOs hosts where the central AGN has been shown to regulate star formation. While evidence for QSO driven winds at low and high-redshifts have been found, only a single case has shown direct evidence that suggests these winds regulate star formation (\citealt{cano-diaz2012}). A larger sample of high-z QSO host galaxy observations are needed to build-up a coherent picture.

We have conducted an IFS LGS-AO pilot survey of five z$\sim$2 type-I QSOs using both Keck II and Gemini North facilities to demonstrate the feasibility and limits of QSO host galaxy detection at high-redshift, and to obtain a range of QSO properties. In \S\ref{obs} we describe observations and target selection. In \S\ref{data} we present the data reduction. In \S\ref{extract} we describe our PSF extraction and removal technique, in \S\ref{results} we discuss our two sources which had a narrow \ha~detection, and describe how we obtained our flux limits in sources with null detections, and in \S\ref{disc} we interpret the results for two of our sources (SDSSJ1029+6510 \& SDSSJ0925+0655) and derive dust-corrected star formation rate limits. We compare our results with studies of QSOs at similar bolometric luminosities, and in \S\ref{conclusion} we provide our conclusions. Throughout the paper we assume a $\Lambda$-dominated cosmology with $\Omega_{M}$=0.308, $\Omega_{\Lambda}$=0.692, and H$_{o}$=67.8 \kms~ Mpc$^{-1}$ \citep{planck14}.

\section{Observations}\label{obs}
We used the near infrared integral field spectrographs OSIRIS (\citealt{larkin2006}) on the Keck telescope and NIFS \citep{mcgregor2003} on the Gemini north telescope (program identification GN-2012B-Q-53) coupled with the observatories' laser guide star adaptive optics systems. We present K-Band spectra of 5 quasars at an average redshift of z $\approx$ 2.2 (angular size scale, 8.5 kpc per arcsecond) with an average total on-target integration time of 3600s. On each night we observed an A type standard star for telluric correction and flux calibration. Table \ref{tabobs} summarizes our observational parameters and setup.

\subsection{Target Selection}\label{targets}
We selected these QSOs from the fifth edition of the SDSS quasar catalog based on the seventh data release (\citealt{schneider2010}). For this pilot survey we selected sources that would have optimal AO performance to aid in the PSF subtraction. Criteria for the Keck and Gemini North observations were: (1) all objects must be observable with the ALTAIR and Keck AO systems based on tip/tilt magnitude and separations (R mag $<$ 16.5 within 25\arcsec~for ALTAIR system and R mag $<$ 18.5 within 45\arcsec~for Keck-AO), and (2) objects must have redshift between 2.016 and 2.427 where H$\alpha$ falls in the prime K-band wavelength regime ($<$ 2.2 $\mu$m). Using these constraints at K-band allowed only $\sim$30 observable QSOs. We made our final selection based on available tip/tilt stars that are bright and close in separation: one with on-axis tip-tilt source correction (R=16.4 mag), and four for off-axis tip-tilt correction. Table \ref{tabobs} contains all the information on the tip/tilt stars. All of our selected sources are Type 1 radio-quiet QSOs with 1.4 GHz flux $<$ 0.15 mJy (\citealt{becker1995}) with no availible near-IR spectroscopy, making our sample less biased towards QSO hosts with high star formation rates. Host galaxies with high star formation rates presented in \citealt{alexander2010}, \citealt{cano-diaz2012} were pre-selected based on long slit spectra of the \oiii5007 \AA~line or far-IR observations.

\subsection{Archival Data}
For multi-wavelength analysis of our objects we include archival observations on our sources. Table \ref{tabphoto} contains optical to near-infrared archival photometric information on our QSO sample, encompassing archival data from the SDSS for the optical magnitudes and 2MASS for near-infrared. As of Data Release 10, SDSS has incorporated WISE and 2MASS photometric data into their catalog, made available in web format on the object explorer website that can be accessed through sdss3.org. In Table \ref{tabphotoIR} we present photometry for the four WISE bands at 3.4\micron, 4.6\micron, 12\micron~and 22\micron. All five sources are detected in the 3.4-12\micron~bands however only three sources have reliable photometry, where the other two suffer from confusion of flux from the bright nearby tip/tilt stars. Three sources are detected in the 22\micron~band, one is undetected and one doesn't have reliable photometry due to confusion; please see Table \ref{tabphotoIR} for details on the individual sources. Two of our sources, SDSSJ1029+651 and SDSSJ2123-005 were observed with the Herschel space telescope's SPIRE instrument\footnote{PIs: D.Weedman, observation ID:1342270222 $\&$ H.Netzer, observation ID:1342270338} in the 250\micron, 350\micron \& 500\micron~bands. We downloaded the fully reduced level 2 maps from the Herschel data archive (http://irsa.ipac.caltech.edu/applications/Herschel/), we converted the maps from Jy/beam to Jy/pixel by dividing the maps by the beam size found in the SPIRE Handbook, available at herschel.esac.esa.int and applied standard aperture photometry over the beam size (17.6\arcsec, 23.9\arcsec, 35.2\arcsec) of the telescope in each of the bands at the optical location of the QSOs from Table \ref{tabobs}. The two sources are undetected in all of the bands and we provide the three sigma limits in Table \ref{tabphotoIR}. 

\section{Data Reduction}\label{data}

\subsection{OSIRIS}\label{osiris}
The OSIRIS observations were reduced using the publicly available OSIRIS data reduction pipeline\footnote{http://www2.keck.hawaii.edu/inst/osiris/tools/}. Dark frames were median combined to produce a master dark frame using the OSIRIS pipeline routine ``combine frames''. Each science and calibration frame then had the master dark subtracted from it and the following pipeline routines were performed: ``adjust channel levels'', ``remove crosstalk'', ``clean cosmic rays'', ``extract spectra'', ''assemble data cube'', ``correct dispersion''. For sky subtraction, each science frame had the nearest in time sky frame subtracted using the ``scaled sky subtraction'' routine that accounts for the temporal variability of the OH sky lines (\citealt{davies2007}). The science and telluric frames were stacked together using a 3$\sigma$ mean clip algorithm in the ``mosaic frames'' routine to remove large bad pixels that occur from the ``extract spectra'' routine . A 1D telluric spectrum was then extracted from the highest signal-to-noise spaxels in the telluric cube using the ''extract star'' routine. Strong hydrogen absorption lines were masked using the ``remove hydrogen lines'' routine, and the blackbody of the star was subtracted using the ``divide blackbody'' routine. The spectra were normalized and used to correct for atmospheric absorption and the instrumental footprint in the mosaiced science frame. The final science data was flux calibrated using standard star observations that were taken closest in time, at similar air mass and were reduced in the same manner as described above.

\subsection{NIFS}\label{nifs}

The NIFS observations were reduced using the Gemini NIFS IRAF reduction pipeline that operates within Pyraf \footnote{http://www.gemini.edu/sciops/instruments/nifs/data-format-and-reduction}. Some modifications were applied to the standard pipeline and additional routines were written to match our science goals. For each night we reduced the Xe, Ar lamp observations to establish the wavelength solution for each of the targets using the Gemini NIFS Pyraf baseline calibration routine. Dark frames for the science observations were median combined and subtracted from each of the science and sky frames. The science, telluric, and sky frames were then reduced using the NIFS science reduction routine. The end result is a data cube which has been flat fielded, bad pixel masked and reformatted into a 3D cube, which was spatially re-sampled from the native spatial sampling of 0.1 $\times$ 0.04\arcsec~to square pixels with a size of 0.05\arcsec. The science and telluric frames had the nearest sky frame in time subtracted, with OH emission line scaling between the sky and science frames. The centroids of the QSO and telluric stars were obtained through a 2D Gaussian fit to a spectrally collapsed image, and the dithered observations were shifted and stacked using a 3$\sigma$ mean clipped algorithm. The 1D telluric star spectrum was extracted by averaging spatially over the highest signal-to-noise spaxels, its blackbody was subtracted, and the strong hydrogen absorption line were masked. The 1D telluric spectrum was then divided into the science cube to correct for atmospheric and instrumental absorption features. 

\section{Extraction of BH masses and PSF subtraction of the QSO}\label{extract}

\begin{figure}[!th]
\includegraphics[width=0.6\textwidth]{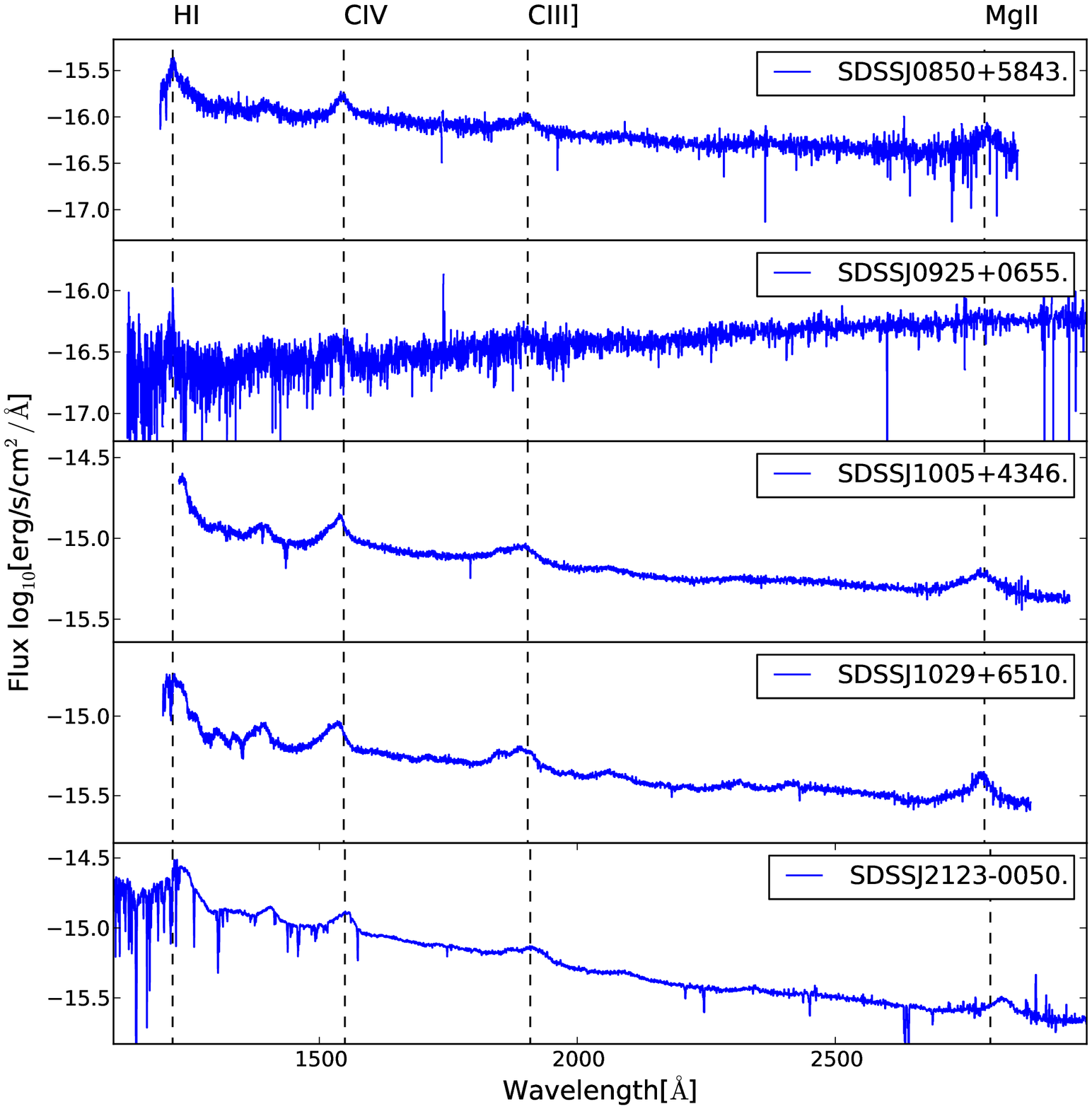}
\includegraphics[width=0.42\textwidth]{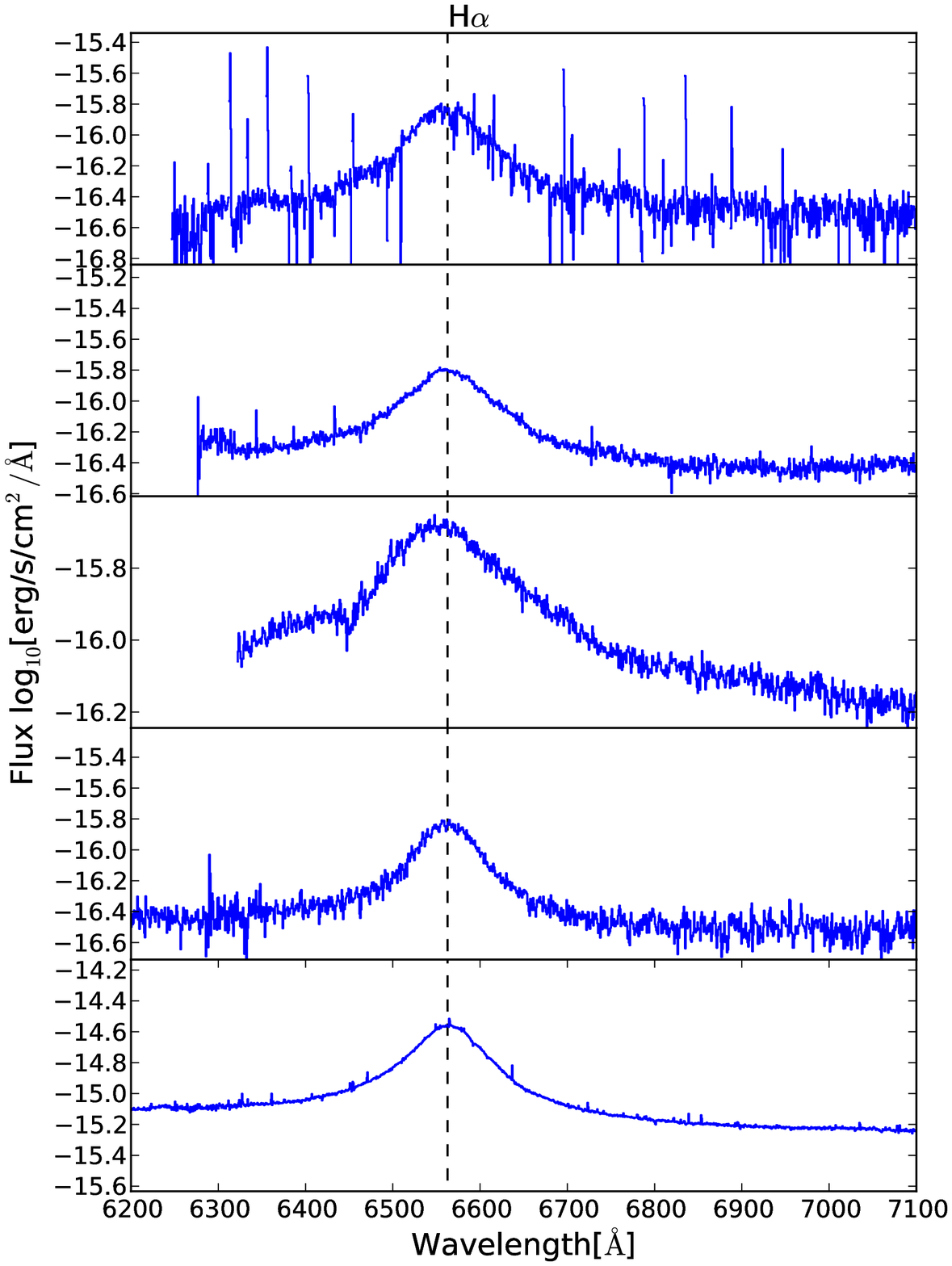}
\caption{SDSS spectra of all the sources in our sample (Left). The SDSS wavelength range covers rest frame UV emission lines of QSOs at this redshift. Vertical dashed lines indicate emission from Ly$\alpha$, CIV, CIII], MgII. Near-IR spectra are presented on the right side, where the broad \ha~line is present. These were extracted from the data cubes, integrating over the spaxels within the seeing halo.}\label{sloan_spec}
\end{figure}

Using the SDSS spectra (Figure \ref{sloan_spec}, left) we derive the bolometric luminosity (L$_{Bol}$) from the rest-frame 1450\AA~continuum using methodology presented in \citealt{runnoe12}. We obtain the black hole mass (M$_{BH}$) using equation (7) presented in \citealt{VP06}, utilizing the 1350\AA~ continuum value together with 1549\AA~CIV FWHM, derived by fitting a Gaussian profile using the curve$\_$fit function that is part of the scipy package, written for python based on non-linear least squares routine. Table \ref{tabprop} contains the above information. Using our K-band QSO spectra (Figure \ref{sloan_spec}, right side) we derive luminosity of the broad \ha~emission line,  black hole mass and the equivalent width (Table \ref{tabpropha}). We fit the line using a Gaussian profile which assisted in deriving the broad \ha~luminosity, redshift and equivalent width. In deriving the line luminosity and equivalent width we integrate over $\pm$1.3$\times$ the FWHM of the fitted profile. The black hole mass was then estimated using equation (6) from \citealt{greene05}. The presented near infrared spectra were extracted from our data cubes using a spatial aperture of approximately the seeing halo.\\
\subsection{PSF construction and subtraction}
The broad \ha~emission originates from gas located in a compact disk within the central few parsecs making this emission essentially point-like in our observations. We use spectral channels that confine the broad line emission for PSF construction. Our algorithm finds the highest signal-to-noise spectral channels that do not coincide with OH emission lines to be combined to create a master PSF image. Generally the selected PSF regions are 2.5-3 nm (10 - 15 spectral channels) in size and tend to sit near the peak of the broad \ha~line. We hypothesised that the majority of the extended narrow line emission will be within 400\kms~from the QSO's redshift, where the PSF has the highest signal to noise and the greatest potential for contamination from the NLR, so we also select spectral regions that are offset from the peak of the broad emission line (not including OH sky lines), that should have minimal contribution from extended narrow-line emission. We combine all spectral regions using a 3$\sigma$ clipping routine, to mitigate contamination from the extended narrow-line emission. This way spaxels that do contain narrow emission would be weighted less since spectral channels offsets by 2000 - 3000\kms~are less likely to contain NLR. The end result is a 2D image of our observed PSF that gets normalized to the flux at the peak pixel. We then go through individual channels in our data cube, scale the image to the maximum value of the PSF at the particular channel and subtract the image. This routine provided the best residuals post PSF subtraction. Some studies have additional steps with PSF construction, by initially fitting and subtracting the nuclear continuum with a low order polynomial (\citealt{inskip11}). The purpose of the linear fit is to remove any continuum emission from the host galaxy. In our work, extensive studies of the final PSF subtracted cube using both methods does not reveal a continuum emission from the host galaxy at the 3$\sigma$ level (average K mag $>$ 20.9), hence we decided not to include this additional step in our QSO PSF construction routine since it adds at least 1.2 times more noise in the PSF subtracted cubes. 

\begin{figure}[!th]
\centering
\includegraphics[width=0.45\textwidth]{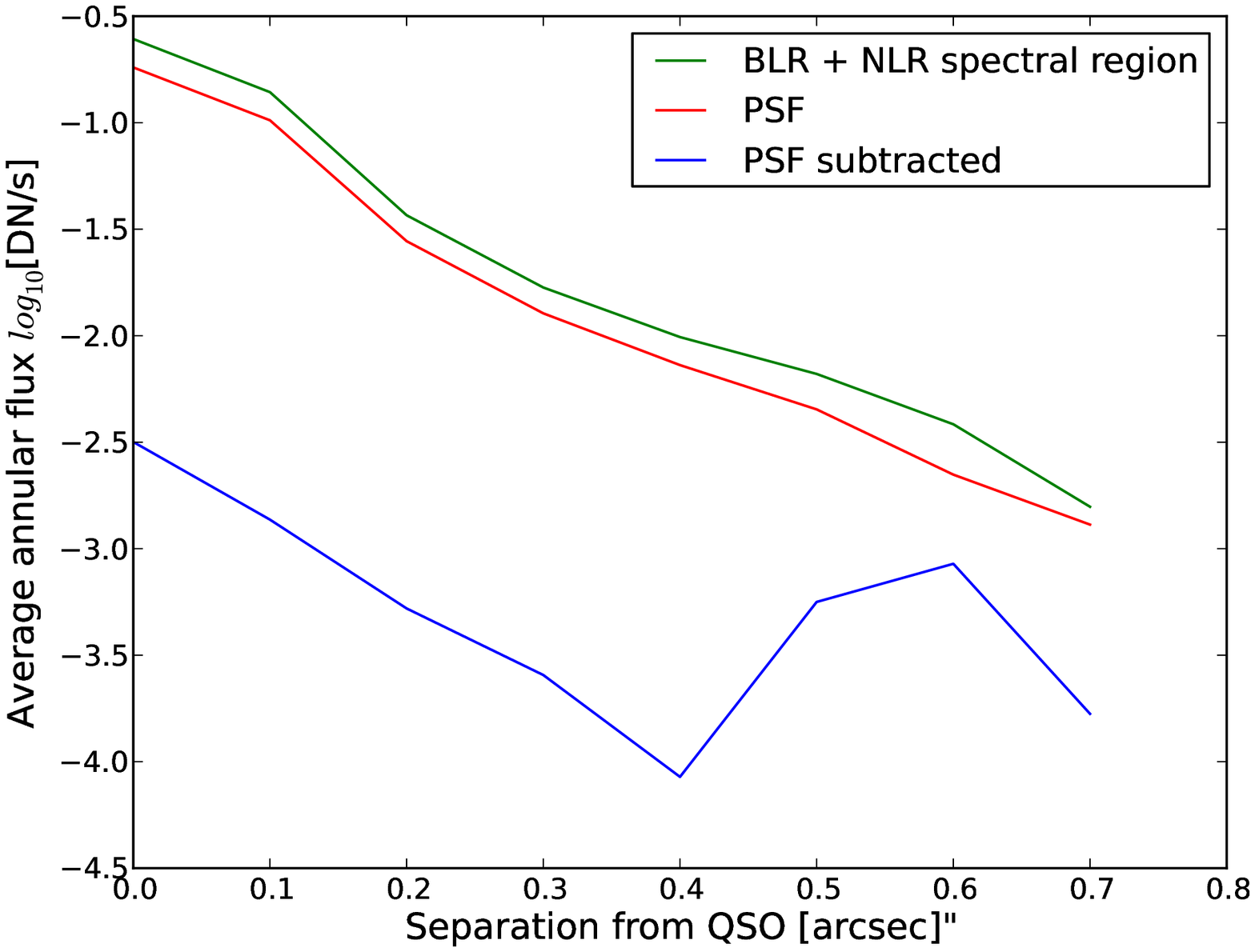}
\includegraphics[width=0.45\textwidth]{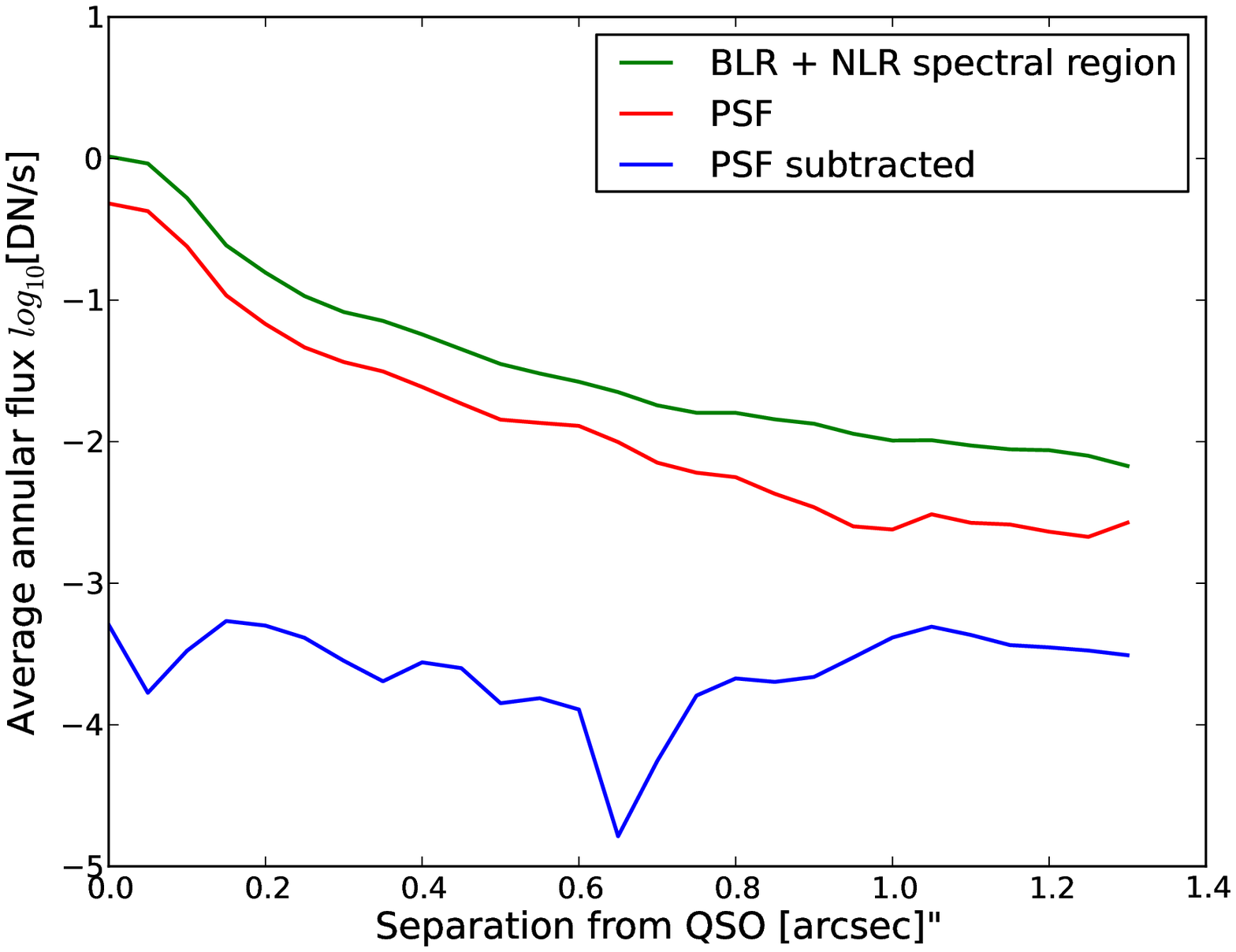}\\
\caption{Radial profiles for SDSSJ1029+6510 (left) and SDSSJ0925$+$0655(right). Green and blue radial profile curves are constructed from spectrally integrated images which contain both broad and narrow \ha, while the red curve is constructed from the PSF image. The blue curve is constructed in the same spectral regions as the green curve however post PSF subtraction, indicating that our PSF removal technique is capable of removing both the AO corrected core as well as the seeing halo.}\label{psfsubresults}
\end{figure}

To test the quality of our PSF subtraction, we constructed radial profiles at different wavelength channels, before and after PSF subtraction, to verify whether the final cube had the central core and seeing halo successfully removed. Figure \ref{psfsubresults} shows the results of these tests for two of our targets. The green and blue radial profile curves are constructed from a spectrally-summed image which contains both broad and narrow \ha~emission ($\Delta\lambda=$2.5 nm or $\Delta$v=1142\kms). The green curve is constructed from the data cube before PSF subtraction, the blue is after the PSF is removed and the red curve is constructed from just the PSF image ($\sim\Delta\lambda=$2.5 nm, spectrally offset 1-5nm). The points are constructed by taking an average in an annulus with $\Delta$r=0.1\arcsec~at a range of separations from the centroid of the QSO. The radial profile in the post PSF subtracted data cube (blue curves) have little slope and significantly less flux, and do not strongly correlate with the general shape of the green and red curves. This demonstrates that the PSF subtracted data has a significant portion of the QSO flux removed, with only the inner 0.2\arcsec~being strongly dominated by noise from PSF subtraction. Averaging over the data cube along the spectral axis, we find that generally within 0.2\arcsec~the QSO still contributes to about 10-20$\%$ of the total data counts, while only 2-5$\%$ outside 0.2\arcsec. As expected, observations with the smallest PSF FWHM showed the best post PSF subtraction data cubes producing the best contrast. However it should be noted that leftover QSO continuum/BLR light does not affect measured values derived from narrow line emission, since they are derived by fitting the line and any underline continuum left over from PSF subtraction simultaneously, at which stage the continuum contamination can be calculated.

\section{Results}\label{results}
To find narrow line emission we searched all of the individual $\sim$ 3,000 spaxels in each of the cubes using an algorithm that searches for flux above a predefined threshold, in combination with visual inspection of each cube. When a line feature is identified we calculated the signal-to-noise by obtaining the standard deviation in the surrounding spatial and spectral pixels, and divided it into the fitted peak of the emission line. For cases where a faint emission feature is found we bin the data using nearby spaxels to increase the signal-to-noise to distinguish between a faint noise spike versus real emission. We confirm a detection if the peak of the emission line is greater than 3$\sigma$ from the neighboring spaxels and the spectral width is greater than the intrinsic instrumental width of 0.35 nm and 0.20 nm for OSIRIS and NIFS respectively. For bright noise spikes we wrote a routine that parses through the cube and removes them if their counts are 5$\sigma$ or higher from the surrounding region (one spaxel in each spatial direction, and 2 spectral channel two the left and right of the spike), some of these features have a FWHM greater than instrumental but given their spatial isolation and significantly higher counts than the surrounding region we quantify them as being "noise spikes". Majority of them are associated with locations of OH sky lines, hence we believe these spikes are residuals caused by sky subtraction. This routine also confirms faint extended structure in the case of SDSSJ0925$+$0655 to be real rather than a combination of separated noise spikes. After searching through the five observed data cubes we identify narrow line emission in two of the systems, SDSSJ1029+6510 and SDSSJ0925+06. For the given QSO redshift, the identified emission lines are likely narrow \ha. If \nii6584\AA~ were assumed instead the flux ratio between it and undetected \ha~would be $\gtrsim$30 in some regions, this is well beyond what has been found in other galaxies (e.g, \citealt{kauffmann2003}). Once an \ha~line is identified we searched for \nii6548,6584 \AA~and \sii6718,6733 \AA~at a similar velocity offset from the broad \ha~line. The detected narrow \ha~emission lines all lie within 600 \kms~from of their respective QSOs broad \ha~redshift, however the full spectral axis in each spaxel was examined for potential narrow emission lines that could be associated with structure surrounding our QSOs. All of the line fits were done with a single Gaussian function using the non-linear least squares routine provided through scipy. The initial guess for the peak is the value at the location of the maximum flux, the initial guess on wavelength offset is the location of the maximum flux, and initial guess on $\sigma$ was 80 \kms, no further constraints were put on the parameters. The radial velocity map is derived from the measured line offset in each spaxel relative to the redshift of the broad \ha~line. The velocity dispersion map of the gas is derived after removing the instrumental width in quadrature from $\sigma$ at each spaxel. Velocity dispersion map is used to dictate the region over which the spectra need to be summed to derive total flux.

\subsection{OSIRIS: SDSSJ1029+6510}\label{osirissdss}
\begin{figure}[!th]
\includegraphics[width=0.45\textwidth]{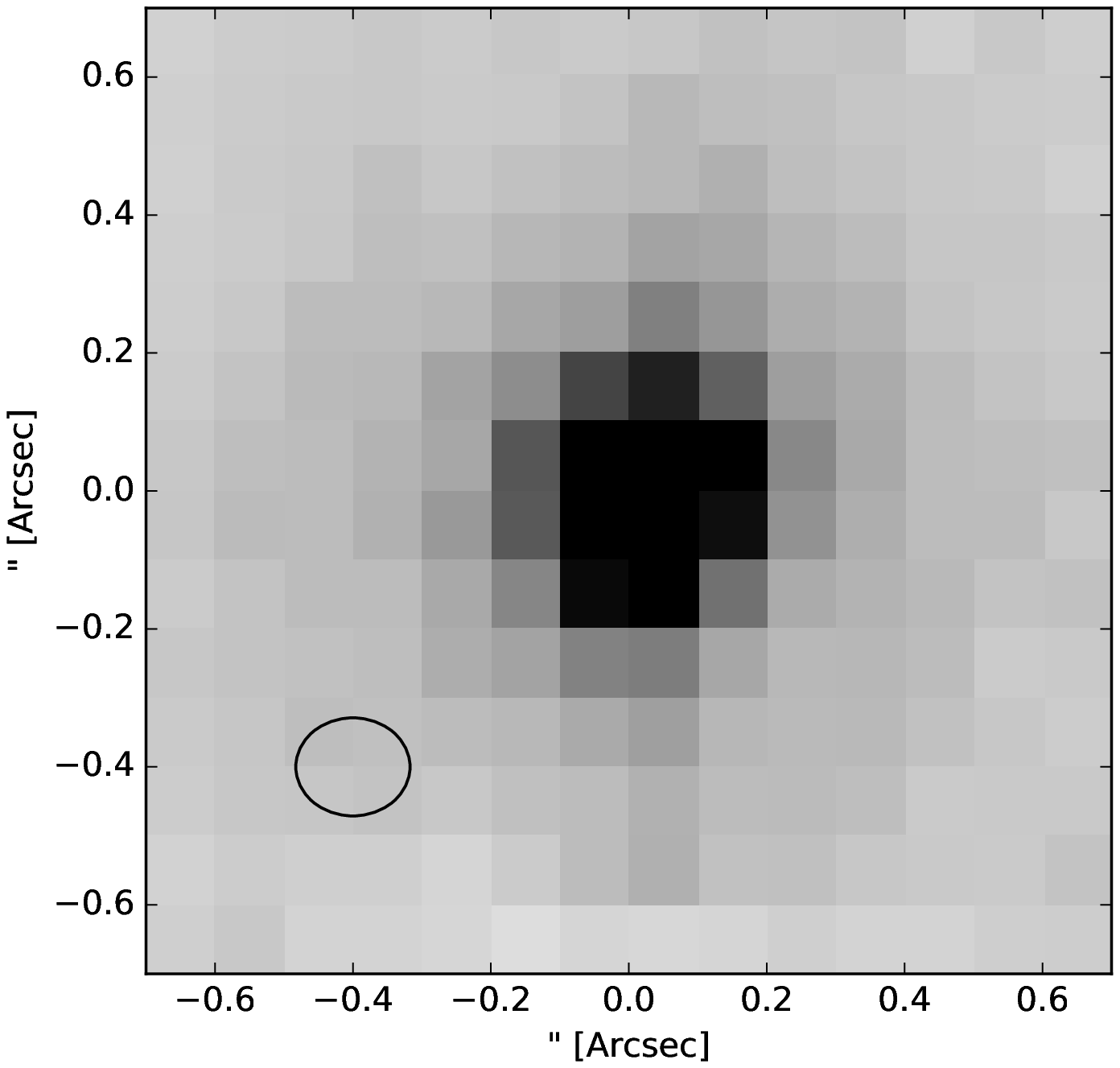}I
\includegraphics[width=0.45\textwidth]{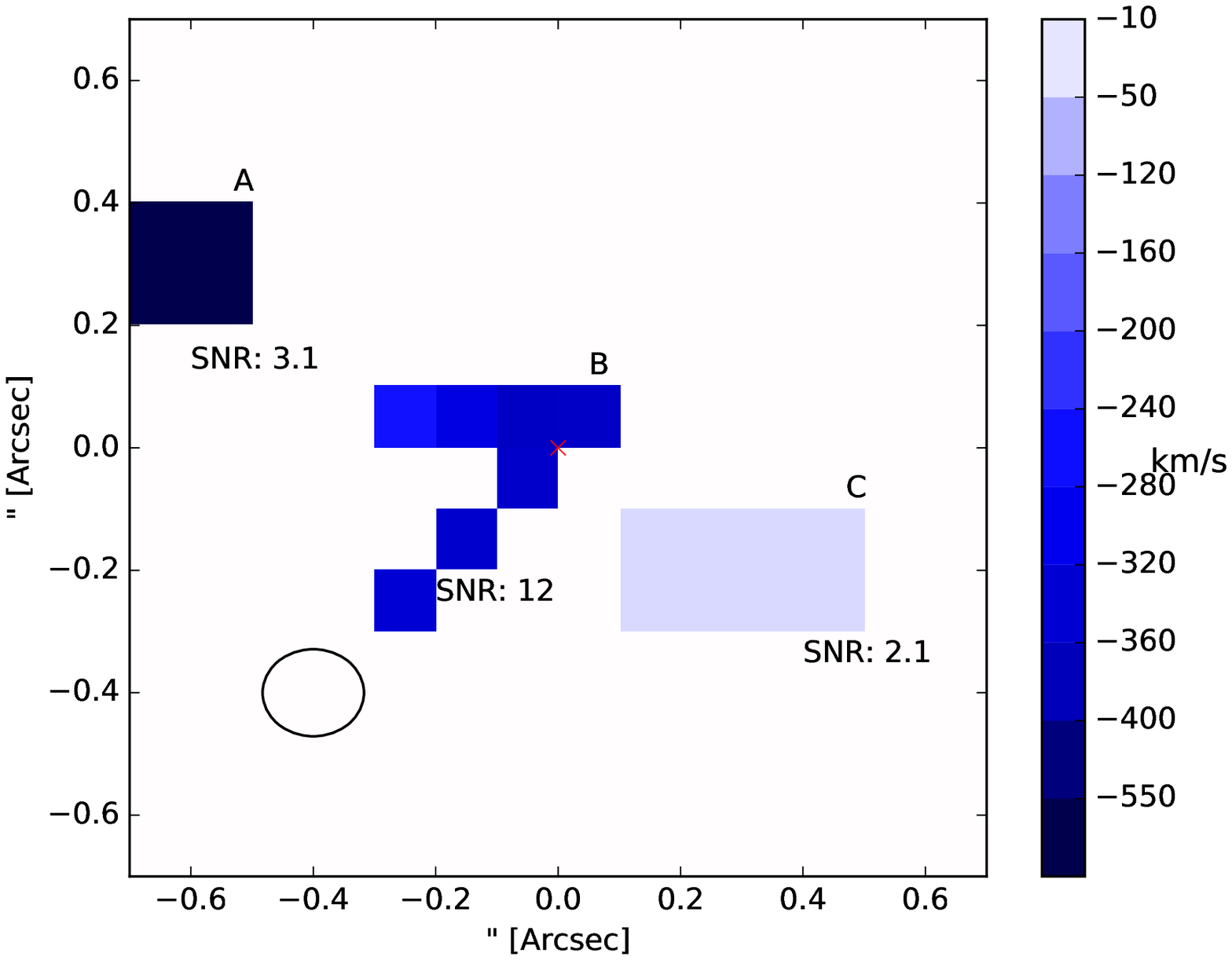}II\\
\includegraphics[width=0.9\textwidth]{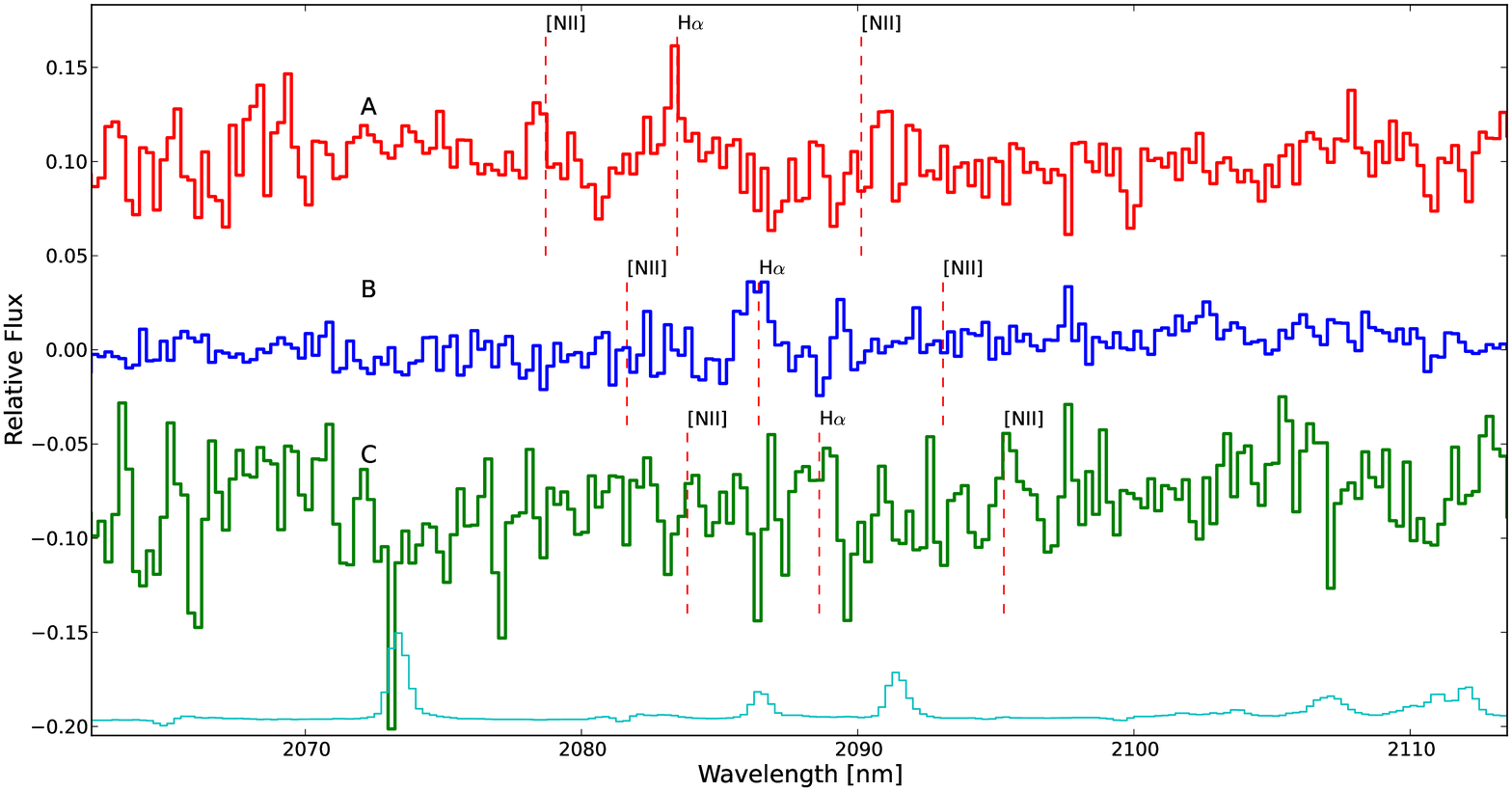}III
\caption{Upper left: K-Band image of SDSSJ1029+6510 from the collapsed OSIRIS LGS-AO data cube using 0.1\arcsec~spatial sampling. Upper right: radial velocity map (\kms) of extended narrow \ha~ emission detected post-PSF subtraction. Radial velocity measurements are obtained by fitting narrow \ha~emission line in the individual regions with a Gaussian function. The spatial resolution of each observation is reprsented by the ellipse in the lower left corner obtained through 2D Gaussian fitting to the PSF image. Bottom: Averaged per spaxel spectra of each of the labeled components with some relative flux offset. The light blue curve shows the wavelength dependence of the noise and OH sky emission. Dashed red lines represent the expected wavelength of narrow emission lines. North is up, east is left.}\label{osiris_results}
\end{figure}

Figure \ref{osiris_results} (panel I) shows the K-band image of the SDSSJ1029+6510 QSO from the collapsed data cube (1.99-2.4 $\micron$). Figure \ref{osiris_results} (panel II) and Figure \ref{osiris_results} (panel III) show the 2D kinematics of the extended narrow line emission relative to the broad \ha~emission and the spectra of the individual components.\\

The PSF subtracted data cube reveals three extended narrow line emission regions, labeled A, B and C in Figure \ref{osiris_results} (panel II). These emission-line regions have a blue-shifted velocity offset of 10-500 \kms~with respect to broad \ha~emission, and a maximum projected separation of $\sim$ 0.6\arcsec (4.2 kpc) from the QSO. We bin the individual spaxels in regions A and C to detect a hint of \ha~emission at a signal-to-noise of 3.1 and 2.1, respectively. Individual spaxels in region B reach a signal-to-noise of $\gtrsim$2, with the central 3 pixels reaching a signal to noise $\gtrsim$ 7. In Table \ref{tabosiris} we present the extracted emission-line properties of the individual regions. Using [N II] $\&$ \ha~we adopt the the line ratio separation between star formation and AGN to be at \loghn=-0.5 in the \hii~diagnostic or ``BPT" \citep{bald81} diagram (Figure \ref{bpt}). The majority of the objects in the region \loghn$<$-0.5 are star forming galaxies (\citealt{kauffmann2003}, \citealt{kewley2001}, \citealt{groves06}). While low metallicity regions ionized by an AGN can be a contaminant at these line ratios, all of the QSOs in our sample (particularly SDSSJ1029+6510) show strong UV emission lines in CIV,SIV+OIV and MgII (Figure \ref{sloan_spec}) that are typical of solar to super-solar metallicity QSOs; hence for this particular system we are not concerned about low metallicity contamination in the region \loghn$<$-0.5. Our limits allow us to discard shock contributions to the emission for regions A and B, line ratios of emission due to shocks tend to reside in \loghn$>$-0.4 on the BPT diagram (\citealt{allen2008}) from a gas that is moving at the recorded velocities of our extended emission. Based on the ratio of \loghn~for A, this region can reside in the transition zone between AGN/SF, assuming no extinction, the star formation rate limit for \ha~flux in region A is 11.0$\pm$2.3 \myr~using the Schmidt-Kennicutt law (SFR$_{H\alpha}=\frac{L_{H\alpha}}{1.26\times10^{41}}$, \citealt{kennicutt98}), this is a limit because AGN photo ionization contribution will increase the observed flux, hence the star formation rate is lower than what is quoted. Region B is located well in the star formation position on the BPT (\loghn$<$-1.5) diagram with a star formation rate limit of 67.4$\pm$5.7\myr. Region C resides well inside the AGN component of the  diagram, and therefore is likely narrow line emission from the QSO, at a projected radial distance of $\sim$2.8 kpc.

\subsection{NIFS: SDSSJ0925$+$0655}\label{nifssdss}
\begin{figure}[!th]
\includegraphics[width=0.45\textwidth]{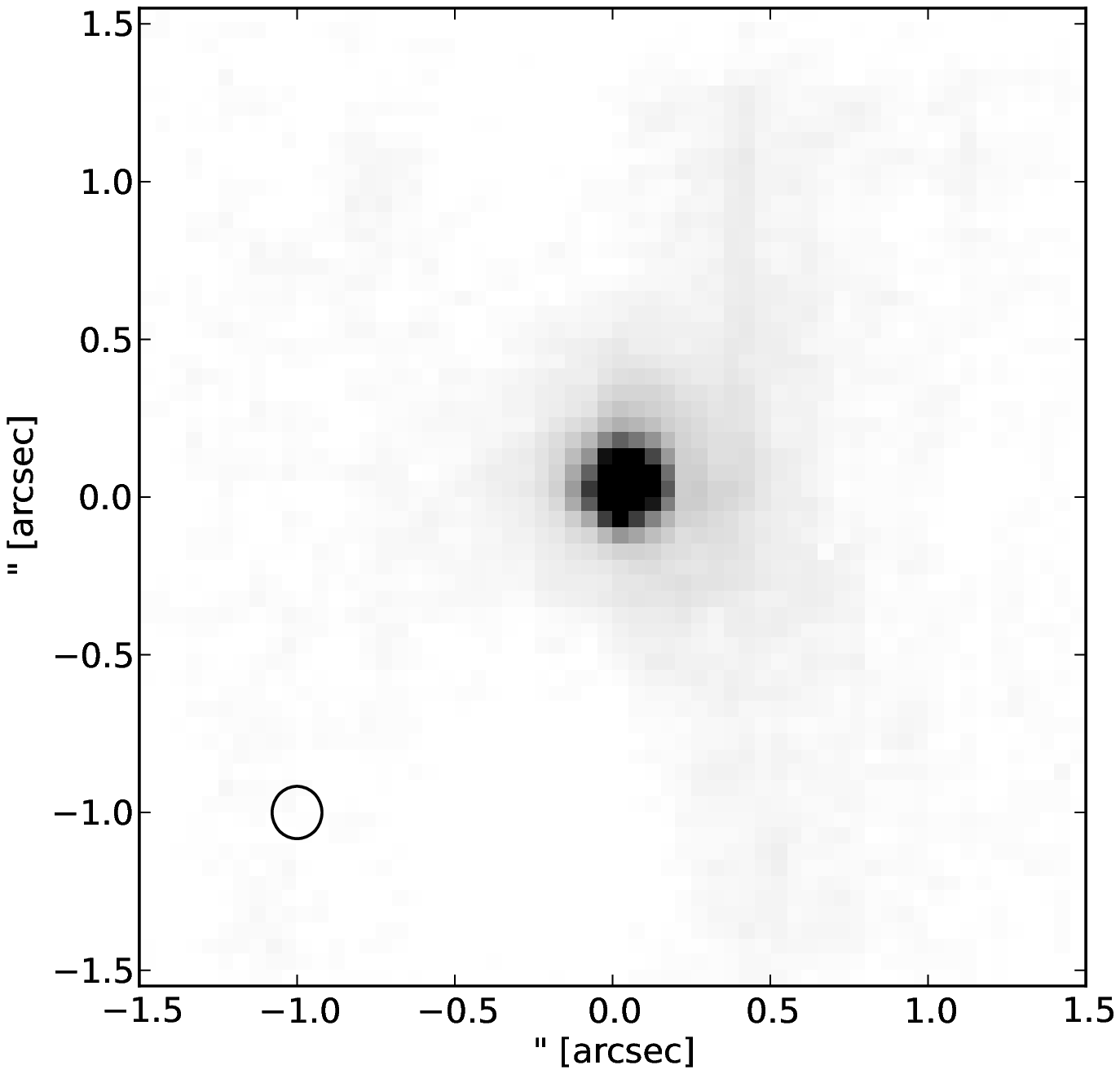}I
\includegraphics[width=0.45\textwidth]{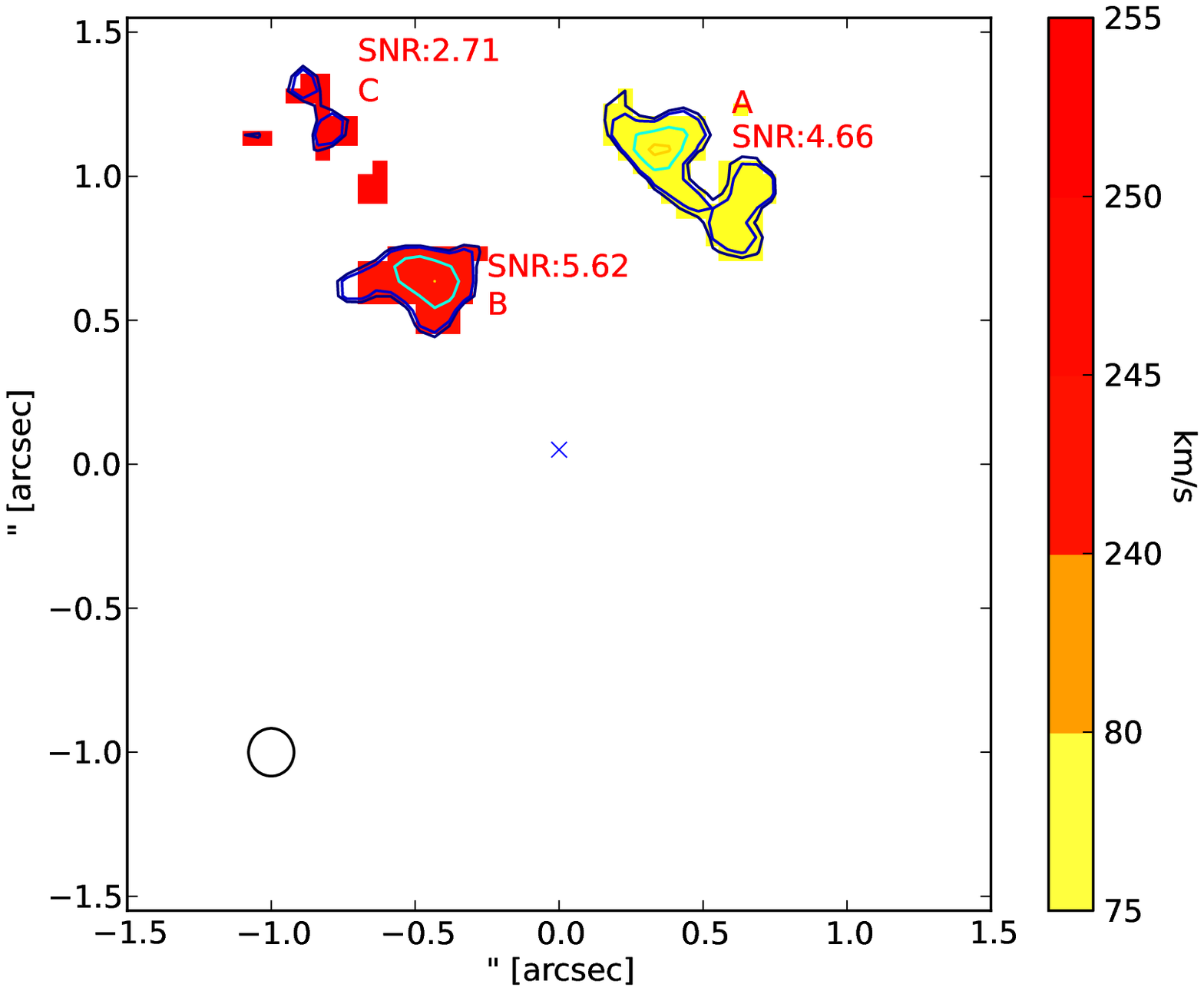}II\\
\includegraphics[width=0.9\textwidth]{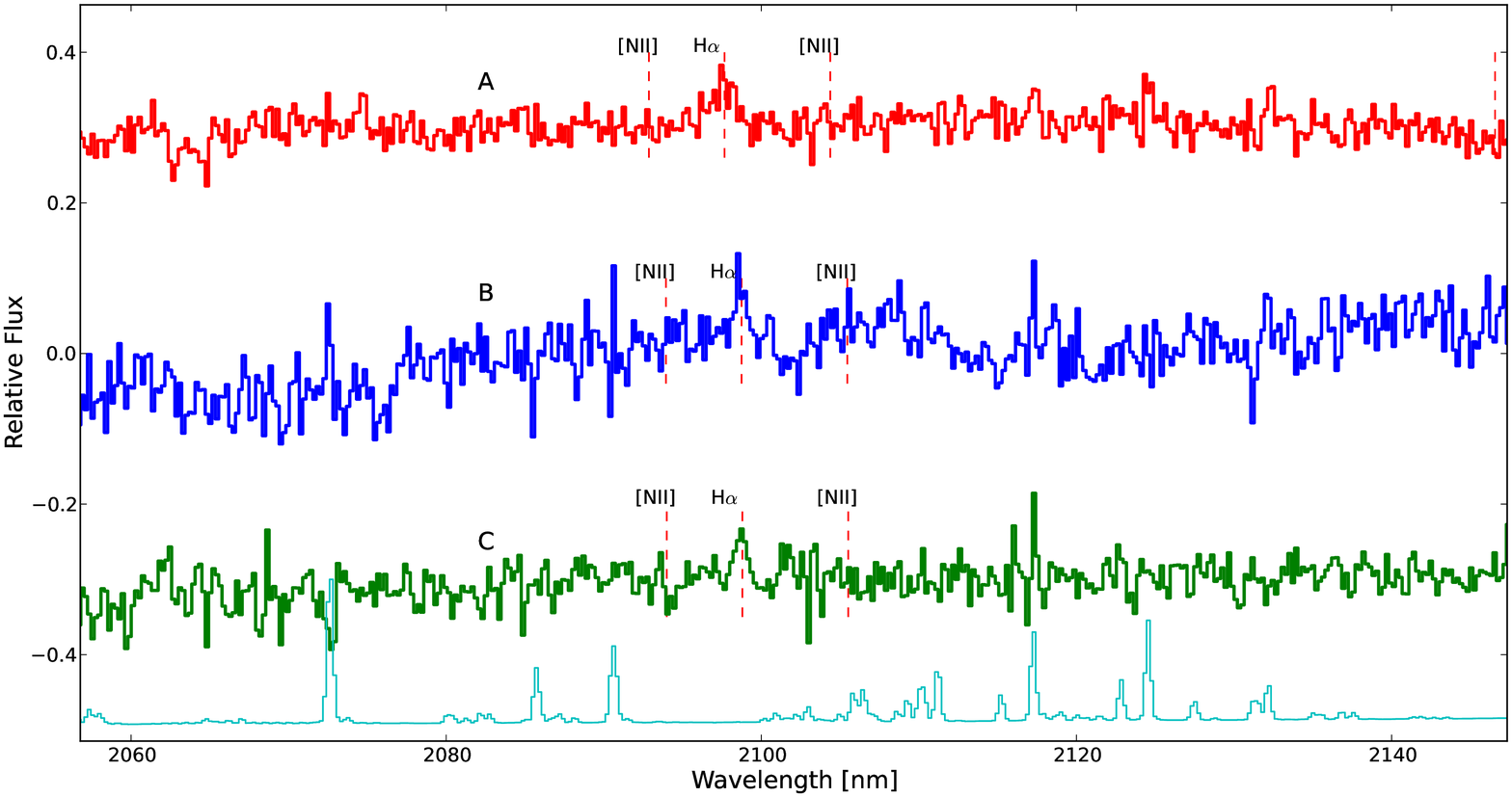}III
\caption{Upper left: K-Band image of SDSSJ0925$+$0655 from the collapsed NIFS Altair AO cube using 0.05\arcsec~spatial sampling. Upper right: PSF subtracted image showing resolved extended \ha~narrow line emission in contours that stretch from 1.8$\sigma$-5$\sigma$ and the velocity map (\kms) obtained from fitting the \ha~line in the individual regions using a Gaussian function. The spatial resolution of each observation is represented by the aperture in the lower left corner obtained through 2D Gaussian fitting to the PSF image. Bottom: Averaged per spaxel spectra of each of the labeled components with some relative flux offset. The light blue curve shows the wavelength dependence of the noise and OH sky. Dashed red lines represent the expected wavelength of narrow emission lines. North is up, east is left.}\label{nifs_results}
\end{figure}

Figure \ref{nifs_results} (panel I) is a K-Band image of the QSO constructed by summing the flux across the entire data cube (1.99-2.4 $\micron$). Figure \ref{nifs_results} (panel II and III), show the 2D kinematics of the extended narrow line emission relative to the redshift of the broad \ha~emission and the spectra of the individual components, respectively. The post-PSF subtracted data cube reveals resolved narrow \ha~emission originating from three distinct regions (A, B, and C), that are both spatially offset (0.5\arcsec-1\arcsec) and redshifted (80-250 \kms) from the QSO, see Table \ref{tabnifs} for extracted parameters on individual regions. We bin by 0.25\arcsec$\times$0.25\arcsec~for each of these regions to increase the signal-to-noise for kinematic analysis. Using ([N II] $\&$ \ha) ratio diagnostic, we put the separation between star formation and AGN at \loghn=-0.5, with star formation being the dominant photoionization mechanism in \loghn$<$-0.5 (see, section \ref{osirissdss} for further discussion). Limits on the \loghn~ratio places regions A,B and C inside the star formation region on the BPT diagram. Our limits allow us to discard shock contributions to the emission for regions A,B, and C, line ratios of emission due to shocks tend to reside in \loghn$>$-0.4 on the BPT diagram (\citealt{allen2008}) from a gas that is moving at the recorded velocities of our extended emission. Using the Schmidt-Kennicutt law (\citealt{kennicutt98}) we obtain un-reddened upper limit star formation rates of 13$\pm$2.3, 12.0$\pm$0.5, 4.0$\pm$0.4 \myr~for regions A, B and C, respectively. Assuming these three clumps have virialized we obtain dynamical masses of 8.7, 1.0, 0.3 $\times10^{9}$M$_{\odot}$ (Table \ref{tabnifs} using the standard virial mass equation $M_{virial}\approx\frac{5R\sigma_{r}^{2}}{G}$)

\subsection{Null detections: SDSSJ1005+4346, SDSSJ2123-0050 \& SDSSJ0850+5843}\label{null}
The remaining three targets reveal no narrow-line \ha~emission offset spatially or spectrally from the QSO. Null detections may be due to two possibilities: (1) these sources have heavy extinction azimuthally around the QSO $\gtrsim$1 kpc; and/or (2) these sources have sufficiently low star-formation rates that reside below the sensitivity limit of these observations.  

We perform a Monte-Carlo simulation in which we generate star forming regions with narrow-line \ha~emission surrounding the QSO at various spatial separations. The purpose of this simulation is to find the limiting flux (and unreddened star-formation rate limits) of our observations and determine how our PSF removal techniques affect our sensitivity versus distance from these QSOs. For our simulations, individual star-forming regions occupy 0.2\arcsec$\times$0.2\arcsec~in the OSIRIS data cube and 0.25\arcsec$\times$0.25\arcsec~in the NIFS data cube, with each spaxel containing a spectrum consisting of an emission line resembling narrow \ha~with a fixed full width at half maximum of 80 \kms (not convolved with an instrumental profile). We select a FWHM of 80 \kms~to match the widths of some of our detected extended narrow line emission, to further test their validity. In a given data cube the star forming regions have a spatially uniform flux, the integrated flux over all the simulated regions vary between cubes. We insert these regions uniformly surrounding the QSO in a cross shape to resemble resolved extended structure, which ranges from 0.1\arcsec~to 1.5\arcsec~in separation from the QSO in the NIFS data cubes and 0.1\arcsec-0.7\arcsec in the OSIRIS cubes. The star forming regions are always centered on the quasar whose position we obtain by fitting a 2D Gaussian to an image of a collapsed data cube along the spectral axis. The spacing between the star forming regions is 0.1\arcsec~to allow signal-to-noise estimates surrounding each individual region. We vary the star formation rates from 0.5 \myr~to 40 \myr~in each of the narrow-line emission regions. For the OSIRIS data, we insert the simulated star forming regions into a data cube that is created by running the extract spectra routine that simply transforms the two dimensional data into a 3D cube. For the Gemini data we run the standard iraf reduction pipeline that extracts the 2D spectra and constructs the 3D data cubes, into which we insert the star forming regions. We then process the data cubes through the rest of the reduction pipeline as described in \S\ref{data}. Finally we run our PSF subtraction routine on the reduced data cubes as described in \S\ref{extract}. We attempt to recover each of the narrow-line \ha~emission regions that were artificially inserted. Just as for the real data, emission must be detected with a minimum of 3$\sigma$ confidence, and emission lines must have a FWMH greater than the instrumental width. 

\begin{figure}[!th]
\centering
\includegraphics[width=0.7\textwidth]{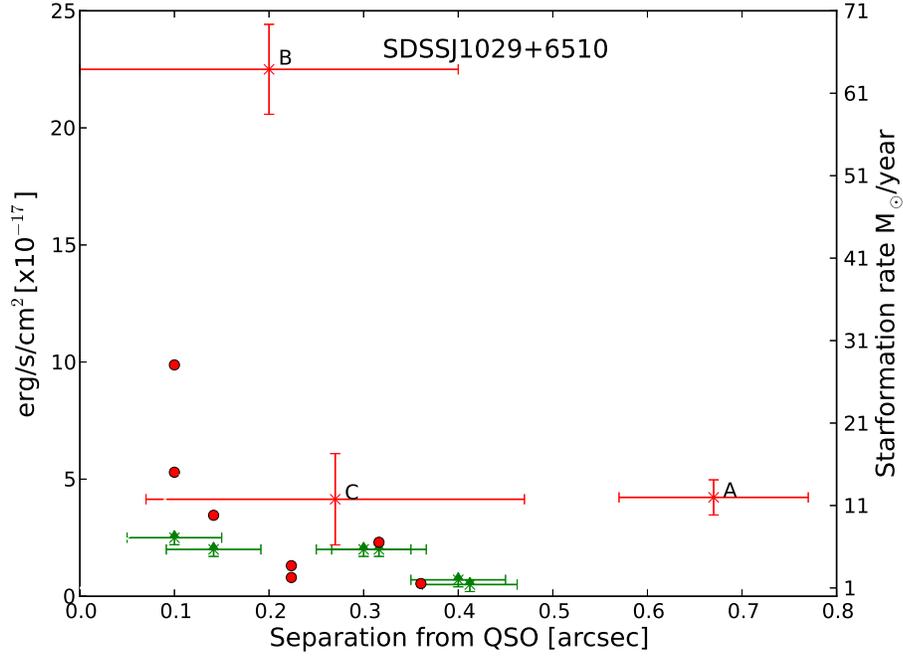}\\
\caption{Limiting integrated flux in a 0.2\arcsec$\times$0.2\arcsec region that was recovered at various separations from the QSO in our Monte Carlo simulation from the OSIRIS observations of SDSSJ1029+6510 (green). Fluxes and distribution of features A, B, C (light red) from Figure \ref{osiris_results} are over plotted. In addition, flux from individual spaxels of region B are plotted in dark red. The three inner spaxels surpass the 0.2\arcsec$\times$0.2\arcsec~box flux limit, while integration of the additional 4 outer spaxels builds a spectrum with a signal-to-noise that is significantly above the noise floor. Note: although the spatial size of region B is greater than 0.2\arcsec$\times$0.2\arcsec, 90\% of the flux sits in the central 3 spaxels, who individually contain a signal to noise ratio $>$ 3.}\label{osiris_limit}
\end{figure}

\begin{figure}[!th]
\centering
\includegraphics[width=0.7\textwidth]{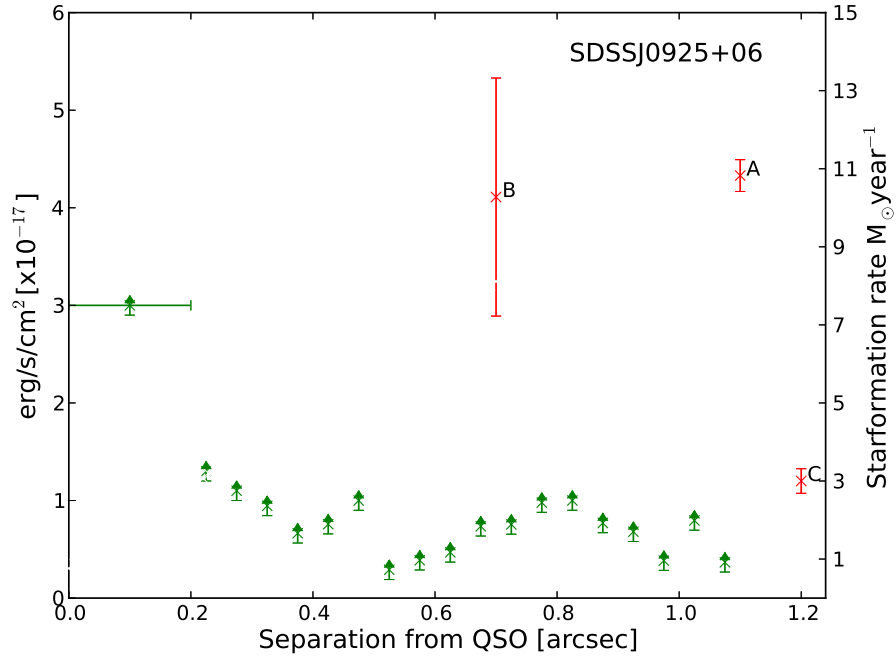}\\
\caption{Limiting integrated flux in a 0.25\arcsec$\times$0.25\arcsec region that was recovered at various separations from the QSO in our Monte Carlo simulations from the NIFS observations of SDSSJ0925+0655 (green). Integrated fluxes of features A, B, C (red) from Figure \ref{nifs_results} are over-plotted.}\label{nifs_limit}
\end{figure}

Recovered star-forming regions with minimum star-formation rates at various angular separations are presented in Figures \ref{osiris_limit} and \ref{nifs_limit}, and fluxes of \ha~ from SDSSJ0925+0655 and SDSSJ1029+6510 regions A,B, and C are over-plotted for comparison. In general we find that our data reduction procedure is not the main factor for missing narrow \ha~flux; the dominant effect is the sensitivity of the detector and PSF removal within 0.2\arcsec~from the QSO. At separations $>0.2$\arcsec~, limiting star-formation rates are an average of 1.4 \myr~(0.7$\times10^{-17}$\ferg) integrated over a star-forming region for the NIFS instrument and 1.5\myr~for OSIRIS. This translates to 0.32\myrkpc~and 0.53\myrkpc~in the NIFS and OSIRIS data cubes respectively.

For SDSSJ1029+6510, we show the integrated flux of region B as well as its individual components in Figure \ref{osiris_limit}, and find they are detected without binning. These simulations and the limiting fluxes for both of these sources indicate low \ha~flux at near and far angular separations from the QSO. For SDSSJ0925+0655, fluxes of the observed components sit well above the star formation distribution (Figure \ref{nifs_limit}), and in principle we are able to detect fainter emission at smaller separations. The other three QSOs do not show any signs of \ha~narrow-line emission.  

We use the bolometric luminosities of our sources to make estimates of dust extinction. The Bolometric luminosities of our sample all sit near 1$\times10^{47}$\ferg, the maximum value for a z$\sim$2 QSO is around 1$\times10^{48}$\ferg~ as has been found by studies such as \citealt{croom09}. This limit only allows us to correct for 2.5 magnitudes of extinction at 1450\AA~, so the limiting star formation rates get as large as 2.03 \myr(0.5\myrkpc) or 2.2 \myr (0.8\myrkpc, using a Small Magellanic cloud extinction curve from \citealt{gordon03}) for NIFS and OSIRIS respectively (see \S\ref{disc1029}, \ref{disc0925}). Note that for SDSSJ0925+0655 the limits may be higher as the QSO is intrinsically redder than the rest of our sample (see \S\ref{disc0925}). We acknowledge that the dust in these scenarios is uniformly distributed, hence the same dust properties that we find along the line of sight to the QSO are elsewhere in the galaxy. Most studies that quote star formation rates give them integrated over some angular scale, typically the beam size of their instrument if the sources they are referencing are not resolved. At the angular resolution of our observations we are capable of resolving a typical z$\sim$2 galaxy with an angular scale of $\sim$1\arcsec. Integrating these limits over a 1\arcsec~ box we obtain for NIFS: 22\myr(33\myr~with maximum dust extinction), OSIRIS: 37\myr (54\myr~with maximum dust extinction). These limits are a sum of the lowest flux that we detected around the QSOs in a 1\arcsec$^{2}$ box in our simulations with the addition of possible dust obscuration. We believe these are hard limits on the upper value of the star formation rate in these host galaxies. Derived SFR limits include contamination from dust in the AGN. There is a possibly that most of the dust is surrounding the nuclear region rather than distributed in the host galaxy. Archival WISE photometry of our sources (Table \ref{tabphotoIR}) shows that 3 of our sources are detected at 22\micron (rest frame $\sim$ 7\micron), all of the sources are detected in the other 3 WISE bands that range from 1-3.75 \micron~at an average redshift of z=2.2, however only 3 sources have reliable photometry due to confusion of flux from the nearby bright tip/tilt stars. For the sources that were detected at an observed wavelength of 22\micron~we find that the average flux density is 16.8\mjy~ indicating that the dust is AGN heated (\citealt{rb95}). Limits closer to the value with minimum dust (22\myr~for NIFS and 37\myr~for OSIRIS) may be more realistic, as some previous studies of dust in type-1 luminous QSOs near z$\sim$2 have found a number of sources with very little (A$_{V}<$0.01) to no extinction (see, \citealt{fynbo2013}). 

\subsection{Unresolved QSO narrow line region emission}
Examining QSO spectra extracted over the PSF halo (Figure \ref{sloan_spec}, right side) we do not detected any unresolved narrow line region emission in any of our sources. We find that generally the spectra are well fitted with a single Gaussian profile and inclusion of narrow emission is only required for the case of SDSSJ1029+65 due to narrow \ha~emission associated with star formation within 0.2\arcsec~of the QSO. We place a flux limit of 3-4$\times10^{-17}$ \ferg~which converts to 1-1.5$\times10^{42}$ \ergs, assuming the NLR emission line has a FWHM of 80 \kms.

\section{Discussion}\label{disc}

There are two explanations for the null narrow-line \ha~emission detections for three of the sources in our sample. This could be caused simply by the lack of star formation and/or significant extinction in the host galaxy. We argue that the main reason we do not see a significant amount of narrow \ha~is likely due to the lack of star formation rather than extinction. Multi-wavelength observations can help estimate the amount of obscuration that is present in the galaxy due to dust. Using available multi-wavelength data we find that our sources do not contain sufficient amounts of dust to cause the observed \ha~limits. The QSOs in our sample are all luminous type-1 AGN representing some of the most powerful QSOs at z$\sim2$. As we will argue in the following sections, even a small dust correction to these systems will increase the bolometric luminosities of our objects above the observed values at this redshift. This indicates that the majority of QSOs in our sample are hosted inside galaxies that are either transitioning from star forming to quenched galaxies or already reside in quiescent galaxies.    

\subsection{SDSSJ1029+6510}\label{disc1029}
\begin{figure}[!th]
\centering
\includegraphics[width=0.6\textwidth]{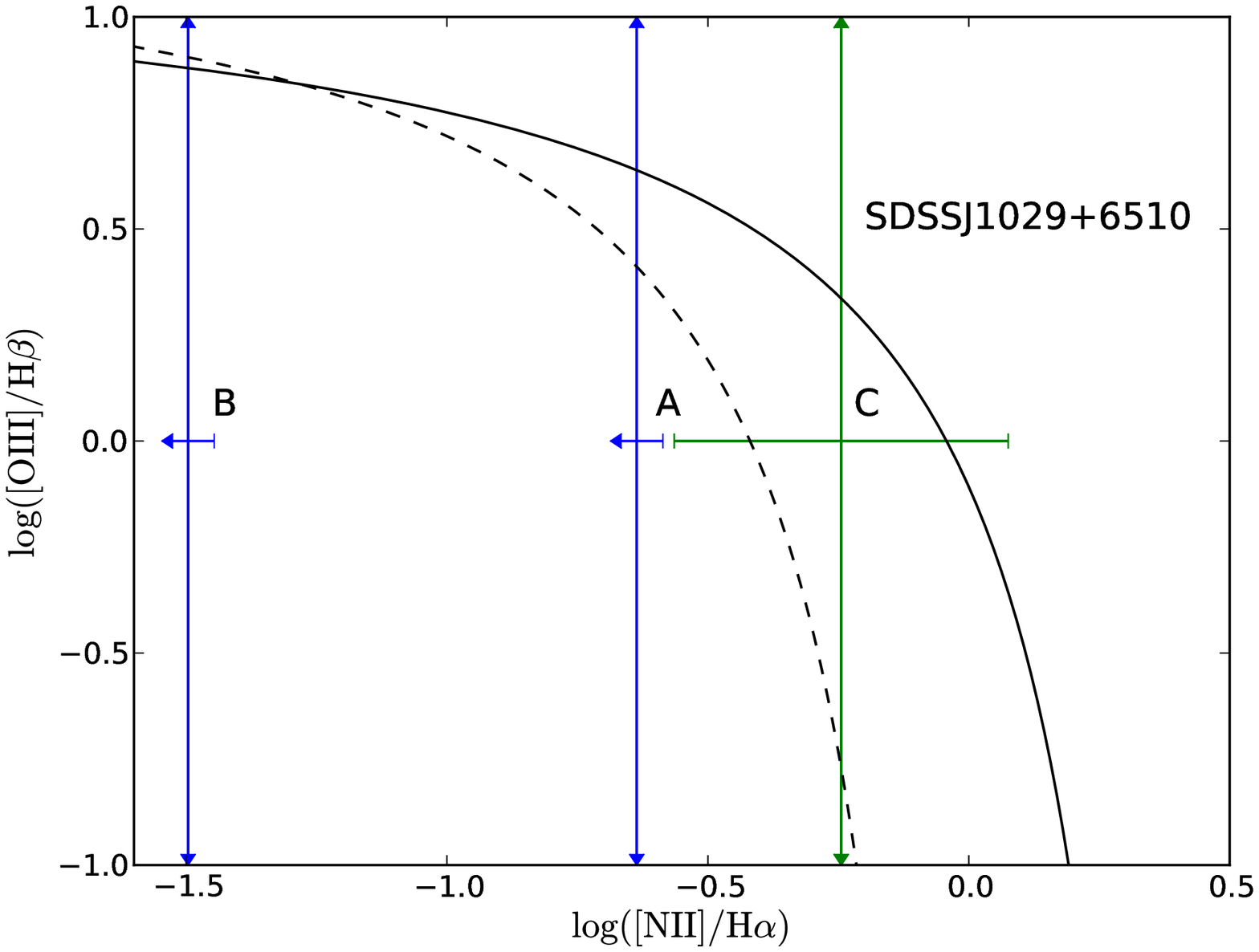}
\includegraphics[width=0.6\textwidth]{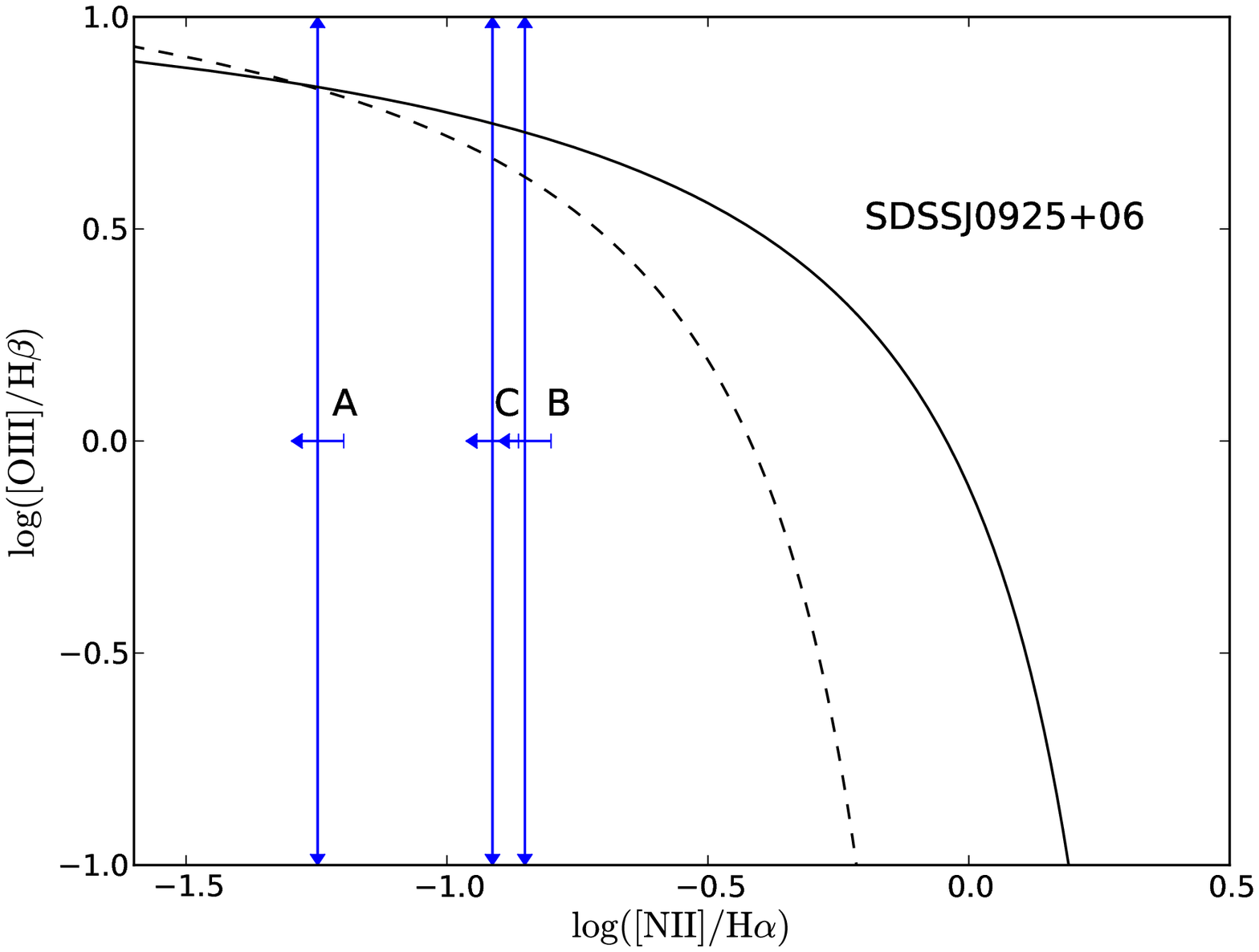}
\caption{Line ratio diagnostics for our detected narrow-line emission using the standard BPT diagram. Limits for log(\oiii/\hb) and log(\nii/\ha) for SDSSJ1029+6510 regions A,B $\&$ C (TOP) and regions A,B, $\&$ C for SDSSJ0925+0655 (BOTTOM) are over-plotted. Dashed curve represents the empirical boundary between star formation and AGN from SDSS (z$<$0.1) by \citealt{kauffmann2003}. The solid curve is the theoretical star formation boundary from \citealt{kewley2001}. Our limits indicate that the narrow-line emission that we detect is not strongly dominated by AGN or shocks, with the possible exception of the low-signal to noise region C in SDSSJ1029+6510.}\label{bpt}
\end{figure}

The host galaxy of this object shows compact vigorous star formation within 2 kpc from the QSO. The rest of the galaxy seems to show no narrow \ha~which we attribute to low star formation rates. SDSSJ1029+6510 is the second most powerful QSO in our sample with a bolometric luminosity of 1.39$\pm0.06\times10^{47}$\ergs (Table-\ref{tabobs}), in addition to the second longest observation time in our sample. Note that some of the emission in individual spaxels of region B are at the 3$\sigma$ level, near the limit of our observations. The ratio of \loghn~$<-1.5$ is located in the \hii~ star formation portion of the diagram (Figure \ref{bpt}) for region B making it a strong candidate for star formation with a formation rate of 67.4$\pm$5.7\myr. This indicates rapid star formation within 2kpc of the QSO.  

For region C, a ratio of 0.57$\pm$0.3 for \loghn~puts this source partly in the AGN ionization region of the diagram (Figure \ref{bpt}), and detection of \nii~emission with higher signal to noise than \ha~suggests this emission is due to the AGN. Lastly for region A, the measured ratio of \loghn=$-0.6$ places it partially inside the star-formation region on the diagram.

This source has a lack of extended star-forming regions, with 90\% of the star formation activity within 2 kpc from the QSO. This is in stark contrast to other resolved host galaxies in \citealt{inskip11}, \citealt{cano-diaz2012} and \citealt{alexander2010}, which have extended star forming regions over several kiloparsecs with star formation rates of $\sim$100 \myr. Our limiting flux simulations indicate that we should detect star formation rates as low as 1.4 \myr~or down to a flux level of 0.6-0.8$\times 10^{-17}$\ferg, at separations $>$0.2\arcsec~from the QSO. Instead, we detect two "streams" (region B at SNR$>$3) of narrow \ha~and nothing else significant around it (regions A and C are $\sim 3\sigma$). This indicates that the surrounding ($>$2 kpc) regions have narrow \ha~flux that is below the sensitivity of the instrument. 

Dust can cause extinction of \ha~flux by re-radiating it at longer wavelength. QSOs in early stages of evolution are thought to be heavily obscured. After the AGN inputs energy/momentum during the ``blow out" phase, gas and dust can get pushed out allowing the AGN \& galaxy to be detected in the optical, which otherwise would be obscured. Observations at other wavelengths can provide clues about the level of obscuration. A strong detection in the far-IR can indicate dust heating due to UV radiation from recent birth of massive stars. This would indicate that some portion of the UV radiation is absorbed (suppressed) and re-emitted at longer wavelength. QSOs that show reddening in their rest-frame UV spectra are good candidates for systems with a considerable level of obscuration, including a number of systems with indicators of outflows through blue shifted broad absorption lines in their rest-frame UV spectra, or broad blue-shifted components in the 500.7nm [OIII] emission line, indicating that some of these systems might be in the ``blow-out" stage (\citealt{farrah2012, urrutia2012}). 

For the case of SDSSJ1029+6510 we are able to put some constraints on the level of obscuration from both far-IR photometry and rest-frame UV-spectrum. This QSO was observed as part of a program with the Herschel space telescope to target some of the brightest optical QSOs with the SPIRE instrument. Examining the archival data we find that at the optical position of the QSO nothing is detected above 3$\sigma$ level in the 250 \micron, 350\micron, and 500\micron~bands. The flux density limits are ($\sim$10mJy, see Table \ref{tabphotoIR}), indicating that this QSO's host galaxy is not in a star-burst phase (L$_{ir}<10^{13}L_{\odot}$). The rest frame UV spectrum obtained from SDSS shows (Figure \ref{sloan_spec}) a continuum slope typical of a type 1 un-obscured QSO (steep blue continuum ), and a bolometric luminosity of 1.39$\times 10^{47}$\ergs~(Table-\ref{tabprop}), which is about an order of magnitude above the average QSO bolometric luminosity. Any correction for dust will start pushing the bolometric luminosity beyond the typical value for bright QSOs at z$\sim2$ ($\sim10^{48}$\ergs). Assuming we need to correct an order of magnitude of flux at rest frame wavelength of 1450\AA~due to dust we would only push the limiting star formation rate to 0.7 \myrkpc~(using a Small Magellanic cloud extinction curve from \citealt{gordon03}), not sufficient to explain the lack of \ha~flux. We therefore favor the low star formation rate model as the main explanation for the observed \ha~flux in the case of SDSSJ1029+6510 at separations greater than 2 kpc.

\subsection{SDSSJ0925+0655}\label{disc0925}
The extended \ha~emission surrounding SDSSJ0925+0655 is a strong candidate for active star formation. The ratio of \loghn~for region A is within the star formation region on the  diagram (Figure \ref{bpt}) while our limits on regions B and C place them near the ambiguous regions between star formation and AGN. The total flux from all these implies an integrated star formation rate of 29$\pm 2.4$ \myr. The detected narrow \ha~emission regions are compact ($\sim$2kpc) and we only detect narrow \ha~in these three regions. In other regions of the data cube we are able to reach a sensitivity limit of 0.8$\times10^{-17}$\ergs~or a star formation rate of 1.4 \myr~at separations $\gtrsim$ 0.2\arcsec~from the QSO. All of the detected regions are at separations $\gtrsim$ 0.5\arcsec~(4 kpc). This implies the narrow \ha~flux sits below the sensitivity of the detector at separations between 1.4 to 4 kpc. We propose that the primary reason for lack of \ha~flux is either from star formation halting, or from obscuration due to dust in the host galaxy (as introduced in the \S\ref{disc1029}). The bolometric luminosity (5.9$\times10^{45}$\ergs) of this QSO as calculated from the 1450\AA~continuum is about an order of magnitude below the average value of a QSO at this redshift, due to the continuum being heavily reddened. However the broad \ha~emission of this source agrees with the rest of the objects in our sample (similar equivalent width and luminosity, see Table \ref{tabprop}) that do not show any signs of reddening in their rest-frame UV spectra (see Figure \ref{sloan_spec} and Table \ref{tabpropha}). The average bolometric luminosities of our sample is $1.24\times10^{47}$\ergs (see Table \ref{tabprop}). The agreement between broad line \ha~properties (velocity dispersion \& intensity) hints that the bolometric luminosity should be consistent with other members of our sample. As found in \citealt{fynbo2013} most reddened QSOs are red due to dust in their host galaxies rather than the inter-galactic medium or dust inside the Milky Way. For this source we estimate the amount of reddening by invoking the condition that the bolometric luminosity should be at the average value for a QSO with such a strong broad \ha~emission (at least $\sim3\times10^{46}$\ergs). This implies that the flux at 1450\AA~ needs to be boosted by $10^{0.94}$ implying that A$_{1450}=2.35$. Using extinction curve from \citealt{gordon03} assuming Small Magellanic Cloud (SMC) like extinction (R$_{v}$=2.74) we obtain A$_{H_{\alpha}}=0.38$. This implies that the flux at \ha~needs to be corrected by at least $10^{0.15}$, yielding a de-reddened star formation rate limit of 0.45 \myrkpc~, and the combined de-reddened star formation rate on A, B, and C of 41 \myr~. This implies that dust attenuation only removes 0.1 \myrkpc~if we only correct the bolometric luminosity such that it sits at the average. Overall this level of dust obscuration is not enough to be the primary reason for low \ha~flux. 

Even assuming an extreme case where the bolometric luminosity is near the maximum value for a type-1 QSO at z$\sim$2 ($\sim10^{48}$) would only imply a limit of 0.9\myrkpc. This could imply that there is a low star formation rate in the host galaxy, where the star formation has been nearly shut off within 0.2\arcsec$-$0.5\arcsec (1.4-4 kpc) from the QSO. These distant regions (A, B and C) are still forming stars at rates that are detectable. Our observations indicate that the host could be in a process of transitioning from a star-forming into a quiescent galaxy. However the less unlikely possibility is that the star formation is active in a diffuse region at separations of 1.4-4kpc rather than in the clumpy regions that we see in regions A, B, C and in other star forming galaxies at this redshift.

\subsection{Comparison to other type-1 QSOs at z$\gtrsim$1}\label{discALMA}
There have been a number of multi-wavelength surveys of radio quiet type 1 QSOs at z$\sim$2 that have presented a range of conclusions about host galaxy star formation properties. High redshift QSO studies have either implied high star formation rates in concurrent high-z type-I QSOs or have argued for a lack of star formation activity. In this section we summarize and compare surveys that share similar QSO properties to our sample (i.e., SMBH mass, bolometric luminosity, unobscured type 1).  

Herschel PACS observations of AGN and QSOs in the COSMOS extragalactic survey indicate a correlation between their bolometric luminosity and rest-frame 60\micron~host galaxy emission (\citealt{rosario2013}). Using the mean 60\micron~flux (3.4$\times10^{45}$\ergs) in the $10^{46-47}$\ergs~z=1.5-2.2 bin in Table 1 from \citealt{rosario2013} indicate that the mean star formation rate should be of order 200 \myr, using the 70\micron~star formation rate law presented in \citealt{calzetti10}. This is nearly an order of magnitude greater than the mean star formation rate in our sample, as indicated by narrow \ha~emission line detection (78\myr~and 29\myr) and limits (22\myr~for NIFS and 37\myr for OSIRIS, integrated over a 1\arcsec$^{2}$ box. See Section \ref{null} for the discussion). The disagreement between our sample and the Herschel results could be due to just the limited-number of sources observed (14 in \citealt{rosario2013} at a similar bolometric luminosity ($10^{45.5-47}$\ergs) as the 5 QSOs in our sample). It is worth noting that the QSOs may be responsible for a significant portion of the total 60\micron~luminosity, so derived 60 \micron~star formation rates should be considered as upper limits. 

HST observations of radio quiet QSOs at z$\sim$2 in \citealt{floyd13} indicate an average star formation rate of 100\myr~derived from rest-frame UV emission originating from the host galaxy. In their study they use both stellar and artificial PSFs to remove the bright QSO. The number of QSOs in our sample is similar to \citealt{floyd13}, which are type-1 and radio quiet. The star formation rate differences between our sample and \citealt{floyd13} could be due to strong QSO contamination from residual emission from their PSF subtraction, or that star formation in our hosts are quite diffuse. 

In contrast, studies such as \citealt{villforth2008} and \citealt{kotilainen2009} find quiescent galaxies that host radio quiet high-z QSOs. These observations are from seeing-limited (0.4-0.5\arcsec) near-infrared imaging and are limited to disentangling the host galaxy at close angular scales ($\lesssim$ 4 kpc). SDSSJ0925+0655 and SDSSJ0850+5843 share similar rest frame UV photometry to their samples, however the other half of the QSOs in our study are 1 to 1.5 magnitudes brighter. Including our results with these two other papers only yields a total of 15 high-z QSO that are observed to reside in ``quiescent" z$\sim$2 galaxies in current literature. 

At even higher redshifts, recent ALMA observations of z$\sim$6 QSOs \citep{wang2013, willott13} using the 158\micron~[CII] emission line reveals a detection in nearly 90$\%$ of the sources observed. The targets in their samples have similar properties to ours (i.e., BH mass, bolometric luminosities $\&$ Eddington ratios). In \citealt{willott13} they reach a star formation limit of 40 \myr~assuming the [CII] emission emanates solely from star formation. Yet sources in \citealt{wang2013} reach star formation rates as high as 1000 \myr, which implies that sources with detected [CII] have extreme star formation rates in comparison to our detections and sensitivity limits at z=2. These z$\sim$6 sources are all near the peak of their starburst phase, assuming that most of the [CII] emission originates from star formation and not the QSO. According to present day M$_{stellar,bulge}$-M$_{bh}$ relation and theoretical work (e.g., \citealt{somerville2008,kor2013}) there is an expectation of simultaneous SMBH and galaxy growth, presumably via mergers at these high ($>1\times10^{46}$\ferg) bolometric luminosities (\citealt{treister12}). In contrast, our observations show star formation rates that are well below this expected initial burst and below the typical star-forming galaxies at z$\sim$2 \citep{erb06,fs09, steidel14}.

The essential difference and advantage of our study compared to previous studies, is that our detection and limits of star formation rates can be made at differing spatial and velocity locations away from the QSO. In contrast, the majority of all studies we have discussed have integrated star formation rate limits over a large range of PSF and beam sizes. Based on our detection limits, it is clear that we do not detect the clumpy (1 kpc$^{2}$), strong star formation regions (up to $\sim$10 \myrkpc) in current IFS observed z$\sim$2 star forming galaxies \citep{fs09,law09,genzel11, law12}. If there is underlying star formation undetected in these host systems, then the surface brightness profiles of the star formation has to be diffuse and integrated across a large area of the galaxy. If our limits are to match previous inferred star formation rates of z$\sim$2 QSO hosts, then it would need to be diffuse with significant extinction.

The sample selection in our pilot survey is albeit random, since we were selecting based on achieving the best AO performance for PSF subtraction, therefore it is interesting that we would happen to select 3/5 type-I QSOs that are quiescent. The majority of our sample is similar to only a small number of observations of high-z QSO hosts residing in quiescent galaxies, and are in disagreement with other work that indicate simultaneous high star formation rates and AGN activity. QSO duty cycles are still poorly understood, however it does seem to appear that in a number of cases the QSO can still be active while star formation in the host has been effectively turned off. These results agree well with AGN feedback models that require that the feedback mechanism only carry a small portion of the total bolometric luminosity of the QSO (5-10$\%$) to effectively turn off star formation \citep{hopkins10}. On the other hand this also agrees with non-causal evolution of SMBH and their host galaxies (\citealt{jm11,peng07}), where the growth of the SMBH and star formation are unrelated and AGN feedback is not the main constituent in formation of local scaling relations, possibly because AGN and star formation activity happen on different time scales. Our study, \citealt{kotilainen2009} and \citealt{villforth2008} are consistent with star formation time scales being significantly shorter than that of the QSO. There are likely numerous high angular resolution observations from HST and ground-based observations that have had null detections of high-redshift QSO host galaxies, that would benefit being released to the community to improve these global statistics. Interestingly, this means there is likely a social selection bias of high-z QSO host galaxies, where authors typically only publish detections (hence QSO hosts with higher star formation properties) rather than their null detections. In any case, it is obvious that there are a large number of selection effects that need to be taken account, but clearly a larger sample of high-redshift QSOs would greatly benefit from IFS+AO observations and aid in our understanding of the demographics of high-z QSO host galaxies.

\section{Conclusions}\label{conclusion}

We have presented LGS-AO assisted integral field spectroscopy observations of five z = 2 QSOs targeted at resolving \ha~nebular emission lines from their host galaxies. Using the broad emission line region of the QSO we were able to construct a PSF to remove the QSO continuum and emission to achieve the necessary contrast to detect \ha~and \nii~host galaxy emission (see \S\ref{extract}). 
\begin{itemize}

\item For two out of five sources (SDSSJ1029+6510 $\&$ SDSSJ0925+0655) we are able to resolve extended narrow line emission surrounding the QSO.

\item In SDSSJ1029+6510 we detect narrow \ha~(regions A and B) that likely originates from star formation at close separations (2 - 4 kpc) from the QSO. If we assume the \ha~flux is from star formation the integrated star formation rate from region A and B is 78.4$\pm$6.2 \myr (110 \myr~with dust correction).

\item For SDSSJ0925+06 we detect three distinct star forming regions that are separated from the QSO by $\sim$ 4 kpc. The upper limit star formation rate for all three regions combined is 29.0$\pm$2.4 \myr (40.7 \myr~with dust corrections).

\item Careful examination of the other three sources in our sample do not detect any narrow \ha~emission post PSF subtraction, even in the cases of SDSSJ1005+4356 $\&$ SDSSJ2123-0050 for which we spent the most integration time per source. 

\item We ran a Monte Carlo simulation on our data by inserting extended narrow \ha~at various separations from the QSO with varying \ha~fluxes (star formation rates). We find that we can detect star formation rates down to 1.4 \myr~(see \S\ref{null}) as close as 0.2\arcsec~from the QSO. Incorporating dust obscuration this value can vary from 2.6\myr-9\myr~ (see \S\ref{disc1029} \& \S\ref{disc0925}) depending on the value of A$_{V}$. At the 9\myr~limit, after correcting the SDSS spectra for dust reddening we are pushing the bolometric luminosities for some of our sources past the typical values for type 1 QSOs at this redshift. Even with a star formation rate of 9 \myr~it would be difficult to explain the missing narrow \ha~to be due to dust obscuration inside the host galaxy. Hence for these sources low star formation rate is the likely reason for lack of narrow \ha~originating from the host galaxy.

\item Four sources show low star formation rates at close angular separation of the QSO, with no dereddened star formation $\gtrsim$ 9 \myr~ within 2 to 4 kpc of the QSO. 

\item We do not detect any strong evidence for NLR emission (region C of SDSSJ1029+6510 is only 2.1$\sigma$) in any of our sources. We place a luminosity limit of 1-1.5$\times10^{42}$\ferg~on an emission line originating from the QSO's NLR. 

\item Compared to other z=2 QSO host galaxy surveys our sample is unique by having little-to-no star formation in high redshift type-I QSOs. This is in agreement with a large fraction of nearby (z $\lesssim$ 0.5) QSO host galaxies being quiescent. Yet at comparable and higher redshifts to our sample the majority of surveys have found simultaneous star formation activity with QSO activity. Clearly a larger z=1-3 QSO IFS+AO sample will be critical in developing a more coherent picture of QSO host galaxies during this important epoch.

\end{itemize}

\acknowledgements
Based on observations obtained at the Gemini Observatory, which is operated by the  Association of Universities for Research in Astronomy, Inc., under a cooperative agreement with the NSF on behalf of the Gemini partnership: the National Science Foundation (United States), the National Research Council (Canada), CONICYT (Chile), the Australian Research Council (Australia), Minist\'{e}rio da Ci\^{e}ncia, Tecnologiae Inova\c{c}\~{a}o (Brazil) and Ministerio de Ciencia, Tecnolog\'{i}a e Innovaci\'{o}n Productiva (Argentina). The authors would like to give our thanks to Eric Steinbring who served as our Gemini phase II liaison for planning this program (GN-2012B-Q-53). Data was also obtained at W.M. Keck Observatory, which was made possible by generous financial support from the W.M. Keck Foundation. The authors would like to acknowledge the dedicated members of the Keck Observatory staff, particularly Jim Lyke and Randy Campbell, who helped with the success of our observations. The authors wish to recognize the significant cultural role and reverence that the summit of Mauna Kea has always had within the indigenous Hawaiian community. We are most fortunate to have the opportunity to conduct observations from this ``heiau" mountain. SCG thanks the Natural Science and Engineering Research Council of Canada for support.



\begin{table}[!th]
\small\addtolength{\tabcolsep}{-4pt}
\begin{center}
\caption{Observational summary of Keck OSIRIS and Gemini ALTAIR-NIFS observations.}\label{tabobs}
\begin{tabular}{lcccccccc}
\tableline\tableline
QSO & R.A & DEC & Observation & Integration &PSF FWHM & Tip/Tilt &Tip/Tilt  \\
SDSSJ  & J2000 & J2000 & Date & N$_{frames}$ $\times$s & \arcsec & sep (\arcsec)&r [mag] \\
\tableline
085022.63+584315.0 &  08:50:22.63   & 58:43:15         &2013Jan04 & 4x600 & 0.15 & 14.1&10.4\\
092547.47+065538.9 & 09:25:47.47    & 6:55:38.9       & 2013Jan08 & 6x600 & 0.13 & 6.2&13.0\\
100517.43+434609.3 & 10:05:17.50 & 43:46:10.9   & 2011Dec31& 16x300 & 0.202 & 21.5& 13.1\\
102907.09+651024.6 & 10:29:07.09 & 65:10:24.6    & 2011Dec30& 12x300 & 0.177 & 43.8&13.8\\
212329.46-005052.9 & 21:23:29.46 & -00:50:52.9      & 2012Aug06& 6x600 & 0.11 & 0&16.4\\
\tableline
\end{tabular}
\end{center}
\end{table}

\begin{table}[!th]
\begin{center}
\caption{SDSS \& 2MASS photometry (\citealt{cutri03}) of the sources in our sample.}\label{tabphoto}
\begin{tabular}{llllllllll}
\tableline\tableline
QSO & u & g& r& i &z&J&H&K\\
\tableline
SDSSJ0850+5843 &19.301 & 19.071& 18.939& 18.977 &18.619&--&--&--&\\
SDSSJ0925+0655 &21.583 & 20.745& 19.888& 19.493 &19.197&--&--&--&\\
SDSSJ1005+4346 &16.985 & 16.803& 16.631& 16.469 &16.249&15.474&15.041&14.271&\\
SDSSJ1029+6510 &17.150& 16.938& 16.833& 16.757&16.602&15.881&15.413&14.566&\\
SDSSJ2123-0050&17.194 &16.648& 16.434& 16.338&16.121&15.180&14.616&13.904&\\
\tableline
\end{tabular}
\end{center}
\end{table}

\begin{table}[!th]
\small\addtolength{\tabcolsep}{-4pt}
\begin{center}
\begin{threeparttable}

\caption{WISE $\&$ Herschel photometry of the five sources in our sample.}\label{tabphotoIR}
\begin{tabular}{lllllllll}
\tableline\tableline
QSO & W1 3.4\micron & W2 4.6\micron& W3 12\micron& W4 22\micron&SPIRE 250\micron&SPIRE 350\micron&SPIRE 500\micron&\\
SDSSJ    &\mjy           &\mjy          &\mjy         &\mjy         &\mjy            &\mjy            &\mjy            &\\
\tableline
0850+5843 &--\tnote{a}&--\tnote{a}&--\tnote{a}&$<$6.8&--&--&--&\\
0925+0655 &--\tnote{a}&--\tnote{a}&--\tnote{a}&--\tnote{a}&--&--&--&\\
1005+4346 &1.32 &2.21& 8.01& 13.83 &--&--&--&\\
1029+6510 &0.87& 1.49& 6.82& 12.91&$<$21.8&$<$13.3&$<$16.5&\\
2123-0050&1.48 &2.37& 9.82& 23.85&$<$13.8&$<$8.67&$<$11.6&\\
\tableline
\end{tabular}
\begin{tablenotes}
\item[a]{No reliable photometry due to source confusion}
\end{tablenotes}
\end{threeparttable}
\end{center}
\end{table}

\begin{table}[!th]
\begin{center}
  \begin{threeparttable}
\caption{QSO General properties \label{tabprop}}
\begin{tabular}{lcccc}
\tableline\tableline
QSO & z$_{UV}$ & L$_{Bol}$ & M$_{BH}$ & Eddington \\
 &    & \ergs$\times10^{47}$ & \msun$\times10^{9}$ & Ratio  \\
\tableline
SDSS0850$+$5843 &2.211 & 0.216$\pm$0.029     & 1.75$\pm$0.13 & 0.098 \\
SDSS0925$+$0655 &2.197 & 0.059$\pm$0.008    & \nodata & \nodata \\
SDSS1005$+$4346 &2.086 & 1.98$\pm$0.06    & 10.2$\pm$0.4 & 0.14 \\
SDSS1029$+$6510 &2.163 & 1.39$\pm$0.06   & 8.0$\pm$0.5 & 0.14 \\
SDSS2123$-$0050 &2.261 & 2.57$\pm$0.07    & 8.5$\pm$0.5 & 0.24 \\
\tableline
\end{tabular}
\begin{tablenotes}
    \item Column 3 is the bolometric luminosities (L$_{Bol}$=4.2$\times L_{1450}$) obtained from rest frame 1450\AA~ continuum with corrections from \citealt{runnoe12}. Column 4 is the black hole mass obtained from rest frame 1549\AA~CIV emission line (\citealt{VP06}). Column 5 is the ratio of the bolometric luminosity to the Eddington luminosity obtained from the measured black hole mass.
\end{tablenotes}
  \end{threeparttable}
\end{center}
\end{table}


\begin{table}[!th]
\begin{center}
  \begin{threeparttable}
\caption{Properties of Broad-line H$\alpha$ emission \label{tabpropha}}
\begin{tabular}{lcccc}
\tableline\tableline
QSO  & z$_{H\alpha}$ & L$_{H\alpha}$ & M$_{BH}$  & Equivalent  \\
 &    & \ergs   & \msun$\times10^{9}$ & Width (\AA) \\
\tableline
SDSS0850$+$5843 &2.212 & 3.16$\times10^{44}$& 1.49$\pm$0.38 & 384$\pm$6 \\
SDSS0925$+$0655 &2.196 &3.58$\times10^{44}$ & 1.94$\pm$0.5 & 352$\pm$1 \\
SDSS1005$+$4346 &2.105 &7.42$\times10^{44}$ & 5.43$\pm$1.47 & 230$\pm$1 \\
SDSS1029$+$6510 &2.183 &3.12$\times10^{44}$ & 0.90$\pm$0.22 & 289$\pm$2 \\
SDSS2123$-$0050 &2.281 &3.84$\times10^{45}$ & 5.0$\pm$1.41 & 281$\pm$1 \\
\tableline
\end{tabular}
\begin{tablenotes}
\item Column 3 is the luminosity of the broad \ha~line. Column 4 is black hole mass derived from \ha~FWHM and its luminosity as in \citealt{greene05}. Column 5 is equivalent width of the broad \ha~line.
\end{tablenotes}
  \end{threeparttable}
\end{center}
\end{table}

\begin{table}[!th]
\addtolength{\tabcolsep}{-4pt}
\begin{center}
  \begin{threeparttable}
\caption{SDSSJ1029+6510: OSIRIS-AO Narrow Emission-line Properties}\label{tabosiris}
\begin{tabular}{cccccccc}
\tableline\tableline
Component & F$_{H\alpha}$ & F$_{[NII]}$6584\AA & [NII]/\ha &  SFR & V$_{r}$ &  V$_{\sigma}$ & M$_{dyn}$ \\
  &  &  &   & \myr &  \kms  &    \kms & \\
\tableline
A&4.22 $\pm$ 0.75&$<$0.951&$<$0.2310 & \nodata &-778$\pm$16 & 163$\pm$36 & \nodata\\
B&22.6 $\pm$ 1.92&$<$0.71&$<$0.0319 & 67$\pm$6 & -355$\pm$19 & 34$\pm$12 & 0.9$\pm$0.07 \\
C&4.14$\pm$ 1.95&2.34$\pm$0.73&-0.24$\pm$ 0.32 & \nodata & -39$\pm$42 & 36$\pm$40 & \nodata\\
\tableline
\end{tabular}
\begin{tablenotes}
\item Column 2 and 3 units are \ferg~$\times$ 10$^{-17}$. Column 8 is in units of M$_{\odot}\times10^{9}$.
\end{tablenotes}
  \end{threeparttable}
\end{center}
\end{table}

\begin{table}[!th]
\addtolength{\tabcolsep}{-4pt}
\begin{center}
  \begin{threeparttable}
\caption{SDSSJ0925+0655: NIFS-AO Narrow Emission-line Properties}\label{tabnifs}
\begin{tabular}{cccccccc}
\tableline\tableline
Component & F$_{H\alpha}$ & F$_{[NII]}$6584\AA & [NII]/\ha &  SFR & V$_{r}$ &  V$_{\sigma}$ & M$_{dyn}$ \\
  &  &  &   & \myr &  \kms  &    \kms & \\
\tableline
A&4.33$\pm$1.22&$<$0.245&$<$0.0565 & 13$\pm$2.3 & 88.4$\pm$19.6 &103.1$\pm$19.3 & 8.7$\pm$4.1 \\
B&4.11$\pm$0.163&$<$0.58&$<$0.1410 & 12$\pm$0.5  &242.6$\pm$15.4 & 37.7$\pm$14.5 & 1.0$\pm$0.8 \\
C&1.20$\pm$0.126&$<$0.148&$<$0.1222 & 4$\pm$0.4  & 250.5$\pm$15.6 & 42.44$\pm$14.7 & 0.3$\pm$0.05 \\
\tableline
\end{tabular}
\begin{tablenotes}
\item Column 2 and 3 units are \ferg~$\times$ 10$^{-17}$. Column 8 is in units of M$_{\odot}\times10^{9}$.
\end{tablenotes}
  \end{threeparttable}
\end{center}
\end{table}

\pagebreak

\begin{thebibliography}{}
\bibitem[Alexander et al.(2010)]{alexander2010} Alexander, D.~M., Swinbank, A.~M., Smail, I., McDermid, R., \& Nesvadba, N.~P.~H.\ 2010, \mnras, 402, 2211 
\bibitem[Alexander \& Hickox(2012)]{alexander2012} Alexander, D.~M., \& Hickox, R.~C.\ 2012, \nar, 56, 93 
\bibitem[Allen et al.(2008)]{allen2008} Allen, M.~G., Groves, B.~A., Dopita, M.~A., Sutherland, R.~S., \& Kewley, L.~J.\ 2008, \apjs, 178, 20
\bibitem[Bahcall et al.(1997)]{Bahcall1997}Bahcall, J.~N., Kirhakos, S., Saxe, D.~H., \& Schneider, D.~P.\ 1997, \apj, 479, 642 
\bibitem[Baldwin et al.(1981)]{bald81} Baldwin, J.~A., 
Phillips, M.~M., \& Terlevich, R.\ 1981, \pasp, 93, 5 
\bibitem[Barai et al.(2014)]{barai2014} Barai, P., Viel, M., Murante, G., Gaspari, M., \& Borgani, S.\ 2014, \mnras, 437, 1456 
\bibitem[Becker et al.(1995)]{becker1995} Becker, R.~H., White, R.~L., \& Helfand, D.~J.\ 1995, \apj, 450, 559
\bibitem[Bennert et al.(2008)]{benn08} Bennert, N., Canalizo, 
G., Jungwiert, B., et al.\ 2008, \apj, 677, 846 
\bibitem[Calzetti et al.(2010)]{calzetti10} Calzetti, D., Wu, S.-Y., Hong, S., et al.\ 2010, \apj, 714, 1256 
\bibitem[Cano-D{\'{\i}}az et al.(2012)]{cano-diaz2012} Cano-D{\'{\i}}az, M., Maiolino, R., Marconi, A., et al.\ 2012, \aap, 537, L8 
\bibitem[Cardelli et al.(1989)]{cardelli89} Cardelli, J.~A., Clayton, G.~C., \& Mathis, J.~S.\ 1989, \apj, 345, 245
\bibitem[Carniani et al.(2013)]{carniani2013} Carniani, S., Marconi, A., Biggs, A., et al.\ 2013, \aap, 559, A29 
\bibitem[Croom et al.(2009)]{croom09} Croom, S.~M., Richards, G.~T., Shanks, T., et al.\ 2009, \mnras, 399, 1755 
\bibitem[Cutri et al.(2003)]{cutri03} Cutri, R.~M., Skrutskie, M.~F., van Dyk, S., et al.\ 2003, ''The IRSA 2MASS All-Sky Point Source Catalog, NASA/IPAC Infrared Science Archive
\bibitem[Davies(2007)]{davies2007} Davies, R.~I.\ 2007, \mnras, 
375, 1099 
\bibitem[Di Matteo et al.(2005)]{dimatteo05} Di Matteo, T., Springel, V., \& Hernquist, L.\ 2005, \nat, 433, 604 
\bibitem[Erb et al.(2006)]{erb06} Erb, D.~K., Steidel, C.~C., Shapley, A.~E., et al.\ 2006, \apj, 647, 128 
\bibitem[Fabian(2012)]{fab12} Fabian, A.~C.\ 2012, \araa, 50, 455
\bibitem[Falomo et al.(2004)]{falomo04} Falomo, R., Kotilainen, J.~K., Pagani, C., Scarpa, R., \& Treves, A.\ 2004, \apj, 604, 495
\bibitem[Falomo et al.(2005)]{falomo05} Falomo, R., Kotilainen, J.~K., Scarpa, ., \& Treves, A.\ 2005, \aap, 434, 469 
\bibitem[Farrah et al.(2012)]{farrah2012} Farrah, D., Urrutia, T., Lacy, M., et al.\ 2012, \apj, 745, 178 
\bibitem[Ferrarese \& Merritt(2000)]{ferra00} Ferrarese, L., \& Merritt, D.\ 2000, \apjl, 539, L9 
\bibitem[Floyd et al.(2010)]{floyd10} Floyd, D.~J.~E., Axon, 
D., Baum, S., et al.\ 2010, \apj, 713, 66 
\bibitem[Floyd et al.(2013)]{floyd13} Floyd, D.~J.~E., Dunlop, 
J.~S., Kukula, M.~J., et al.\ 2013, \mnras, 429, 2
\bibitem[F{\"o}rster Schreiber et al.(2009)]{fs09} F{\"o}rster Schreiber, N.~M., Genzel, R., Bouch{\'e}, N., et al.\ 2009, 
\apj, 706, 1364
\bibitem[Fynbo et al.(2013)]{fynbo2013} Fynbo, J.~P.~U., Krogager, J.-K., Venemans, B., et al.\ 2013, \apjs, 204, 6 

\bibitem[Gebhardt et al.(2000)]{geb00} Gebhardt, K., Bender, R., Bower, G., et al.\ 2000, \apjl, 539, L13 
\bibitem[Genzel et al.(2011)]{genzel11} Genzel, R., Newman, S., 
Jones, T., et al.\ 2011, \apj, 733, 101 
\bibitem[Gordon et al.(2003)]{gordon03} Gordon, K.~D., Clayton, G.~C., Misselt, K.~A., Landolt, A.~U., \& Wolff, M.~J.\ 2003, \apj, 594, 279 
\bibitem[Greene \& Ho(2005)]{greene05} Greene, J.~E., \& Ho, L.~C.\ 2005, \apj, 630, 122
\bibitem[Groves et al.(2006)]{groves06} Groves, B.~A., Heckman, T.~M., \& Kauffmann, G.\ 2006, \mnras, 371, 1559
\bibitem[Harrison et al.(2014)]{harrison2014} Harrison, C.~M., Alexander, D.~M., Mullaney, J.~R., \& Swinbank, A.~M.\ 2014, arXiv:1403.3086 
\bibitem[Hopkins \& Quataert(2011)]{hopkins11} Hopkins, P.~F., \& Quataert, E.\ 2011, \mnras, 415, 1027 
\bibitem[Hopkins \& Elvis(2010)]{hopkins10} Hopkins, P.~F., \& Elvis, M.\ 2010, \mnras, 401, 7
\bibitem[Jahnke \& Macci{\`o}(2011)]{jm11} Jahnke, K., \& Macci{\`o}, A.~V.\ 2011, \apj, 734, 92 
\bibitem[Liu et al.(2013)]{liu13} Liu, G., Zakamska, N.~L., Greene, J.~E., Nesvadba, N.~P.~H., \& Liu, X.\ 2013, \mnras, 436, 2576
\bibitem[Hamilton et al.(2002)]{ham02} Hamilton, T.~S., 
Casertano, S., \& Turnshek, D.~A.\ 2002, \apj, 576, 61 
\bibitem[Hutchings et al.(2002)]{hutching2002} Hutchings, J.~B., Frenette, D., Hanisch, R., et al.\ 2002, \aj, 123, 2936 
\bibitem[Inskip et al.(2011)]{inskip11} Inskip, K.~J., Jahnke, K., Rix, H.-W., \& van de Ven, G.\ 2011, \apj, 739, 90
\bibitem[Kornei et al.(2012)]{kornei2012} Kornei, K.~A.,Shapley, A.~E., Martin, C.~L., et al.\ 2012, \apj, 758, 135
\bibitem[Kotilainen et al.(2009)]{kotilainen2009} Kotilainen, J.~K., Falomo, R., Decarli, R., et al.\ 2009, \apj, 703, 1663
\bibitem[Kauffmann et al.(2003)]{kauffmann2003} Kauffmann, G., Heckman, T.~M., Tremonti, C., et al.\ 2003, \mnras, 346, 1055 
\bibitem[Kennicutt(1998)]{kennicutt98} Kennicutt, R.~C., Jr.\ 1998, \apj, 498, 541
\bibitem[Kewley et al.(2001)]{kewley2001} Kewley, L.~J., Dopita, M.~A., Sutherland, R.~S., Heisler, C.~A., \& Trevena, J.\ 2001, \apj, 556, 121 
\bibitem[Kirhakos et al.(1999)]{kirhakos1999} Kirhakos, S., Bahcall, J.~N., Schneider, D.~P., \& Kristian, J.\ 1999, \apj, 520, 67 
\bibitem[Kormendy \& Ho(2013)]{kor2013} Kormendy, J., \& Ho, L.~C.\ 2013, \araa, 51, 511
\bibitem[Larkin et al.(2006)]{larkin2006} Larkin, J., Barczys, M., Krabbe, A., et al.\ 2006, \nar, 50, 362 
\bibitem[Law et al.(2009)]{law09} Law, D.~R., Steidel, C.~C., 
Erb, D.~K., et al.\ 2009, \apj, 697, 2057 
\bibitem[Law et al.(2012)]{law12} Law, D.~R., Shapley, A.~E., 
Steidel, C.~C., et al.\ 2012, \nat, 487, 338 
\bibitem[Lehnert et al.(1999)]{lehnhert99} Lehnert, M.~D., van Breugel, W.~J.~M., Heckman, T.~M., \& Miley, G.~K.\ 1999, \apjs, 124, 11
\bibitem[Magorrian et al.(1998)]{mago1998} Magorrian, J., Tremaine, S., Richstone, D., et al.\ 1998, \aj, 115, 2285 
\bibitem[M{\'a}rquez \& Petitjean(2003)]{mar03} M{\'a}rquez, I., \& Petitjean, P.\ 2003, Revista Mexicana de Astronomia y Astrofisica Conference Series, 16, 135
\bibitem[Matsuoka et al.(2014)]{mats14} Matsuoka, Y., Strauss, 
M.~A., Price, T.~N., III, \& DiDonato, M.~S.\ 2014, \apj, 780, 162 
\bibitem[McConnell \& Ma(2013)]{2013ApJ...764..184M} McConnell, N.~J., \& Ma, C.-P.\ 2013, \apj, 764, 184 
\bibitem[McGregor et al.(2003)]{mcgregor2003} McGregor, P.~J., Hart, J., Conroy, P.~G., et al.\ 2003, \procspie, 4841, 1581 
\bibitem[O'Donnell(1994)]{odonnell94} O'Donnell, J.~E.\ 1994, \apj, 422, 158
\bibitem[Peng(2007)]{peng07} Peng, C.~Y.\ 2007, \apj, 671, 1098 
\bibitem[Pettini \& Pagel(2004)]{pettini2004} Pettini, M., \& Pagel, B.~E.~J.\ 2004, \mnras, 348, L59 
\bibitem[Planck Collaboration et al.(2014)]{planck14} Planck Collaboration, Ade, P.~A.~R., Aghanim, N., et al.\ 2014, \aap, 571, AA16 
\bibitem[Quinlan \& Hernquist(1997)]{QH2001} Quinlan, G.~D., \& Hernquist, L.\ 1997, \na, 2, 533 
\bibitem[Somerville et al.(2008)]{somerville2008} Somerville, R.~S., 
Hopkins, P.~F., Cox, T.~J., Robertson, B.~E., 
\& Hernquist, L.\ 2008, \mnras, 391, 481
\bibitem[Ridgway et al.(2002)]{ridgeway02} Ridgway, S., Heckman, 
T., Calzetti, D., \& Lehnert, M.\ 2002, \nar, 46, 175 
\bibitem[Riechers et al.(2008)]{riechers2008} Riechers, D.~A., Walter, F., Carilli, C.~L., Bertoldi, F., \& Momjian, E.\ 2008, \apjl, 686, L9 
\bibitem[Rosario et al.(2013)]{rosario2013} Rosario, D.~J., Trakhtenbrot, B., Lutz, D., et al.\ 2013, \aap, 560, A72 
\bibitem[Rowan-Robinson(1995)]{rb95} Rowan-Robinson, M.\ 1995, \mnras, 272, 737 
\bibitem[Runnoe et al.(2012)]{runnoe12} Runnoe, J.~C., Brotherton, M.~S., \& Shang, Z.\ 2012, \mnras, 422, 478 
\bibitem[Sesana et al.(2008)]{sesana2008} Sesana, A., Haardt, F., \& Madau, P.\ 2008, \apj, 686, 432 
\bibitem[Scannapieco et al.(2005)]{scan05} Scannapieco, E., 
Silk, J., \& Bouwens, R.\ 2005, \apjl, 635, L13
\bibitem[Schaerer et al.(2014)]{schaerer2014} Schaerer, D., Boone, F., Zamojski, M., et al.\ 2014, arXiv:1407.5793
\bibitem[Schneider et al.(2010)]{schneider2010} Schneider, D.~P., Richards, G.~T., Hall, P.~B., et al.\ 2010, \aj, 139, 2360 
\bibitem[Schramm et al.(2008)]{schramm2008} Schramm, M., Wisotzki, L., \& Jahnke, K.\ 2008, \aap, 478, 311 
\bibitem[Steidel et al.(2014)]{steidel14} Steidel, C.~C., Rudie, 
G.~C., Strom, A.~L., et al.\ 2014, arXiv:1405.5473 
\bibitem[Thompson et al.(2005)]{thompson2005}{2005ApJ...630..167T} Thompson, T.~A., Quataert, E., \& Murray, N.\ 2005, \apj, 630, 167 
\bibitem[Treister et al.(2012)]{treister12} Treister, E., Schawinski, K., Urry, C.~M., \& Simmons, B.~D.\ 2012, \apjl, 758, L39 
\bibitem[Urrutia et al.(2012)]{urrutia2012} Urrutia, T., Lacy, M., Spoon, H., et al.\ 2012, \apj, 757, 125
\bibitem[Veilleux et al.(2009)]{veilleux09} Veilleux, S., Kim, 
D.-C., Rupke, D.~S.~N., et al.\ 2009, \apj, 701, 587 
\bibitem[Vestergaard \& Peterson(2006)]{VP06} Vestergaard, M., \& Peterson, B.~M.\ 2006, \apj, 641, 689 
\bibitem[Villforth et al.(2008)]{villforth2008} Villforth, C., Heidt, J., \& Nilsson, K.\ 2008, \aap, 488, 133 
\bibitem[Wang et al.(2013)]{wang2013} Wang, R., Wagg, J., Carilli, C.~L., et al.\ 2013, \apj, 773, 44 
\bibitem[Wang et al.(2009)]{wang09} Wang, J.-G., Dong, X.-B., Wang, T.-G., et al.\ 2009, \apj, 707, 1334
\bibitem[Willott et al.(2013)]{willott13} Willott, C.~J., Omont, A., \& Bergeron, J.\ 2013, \apj, 770, 13 
\bibitem[Wurster \& Thacker(2013)]{wurster13} Wurster, J., \& Thacker, R.~J.\ 2013, \mnras, 431, 2513 
\bibitem[Zakamska et al.(2006)]{zak06} Zakamska, N.~L., 
Strauss, M.~A., Krolik, J.~H., et al.\ 2006, \aj, 132, 1496 



\end{thebibliography}
\end{document}